\documentclass[fleqn,usenatbib]{mnras}

\usepackage[T1]{fontenc}
\usepackage{ulem}
\DeclareRobustCommand{\VAN}[3]{#2}
\let\VANthebibliography\thebibliography
\def\thebibliography{\DeclareRobustCommand{\VAN}[3]{##3}\VANthebibliography}

\usepackage{graphicx}	
\usepackage{amsmath}	
\usepackage{array}
\usepackage{mathabx}
\usepackage{soul}
\usepackage{adjustbox}
\usepackage{subcaption}

\title{The Orbit of WASP-4\,b is in Decay}
\author[Ba\c{s}t\"urk et al.]{%
\"O.\ Ba\c{s}t\"urk$^{1,2}$\thanks{E-mail: obasturk@ankara.edu.tr},
A.\ C.\ Kutluay$^{2,3}$,
A.\ Barker$^{4}$,
S.\ Yal\c{c}{\i}nkaya$^{1,2,3}$,
J.\ Southworth$^{5}$,
K.\ Barkaoui$^{6,7,8}$,
\newauthor
A.\ W\"unsche$^{9}$,
M.\ J.\ Burgdorf$^{10}$,
M.\ Timmermans$^{6,11}$,
E.\ Jehin$^{12}$,
J.\ Tregloan-Reed$^{13}$,
\newauthor
R.\ Figuera Jaimes$^{14,15,16}$,
T.\ C.\ Hinse$^{17}$, 
B.\ Duru$^{2,3}$,
J.\ Hitchcock$^{16}$,
P.\ Longa-Pe{\~n}a$^{18}$,
S.\ Rahvar$^{19}$,
\newauthor
S.\ Sajadian$^{19}$,
M.\ Bretton$^{9}$,
S.\ O.\ Selam$^{1,2}$,
M.\ Gillon$^{6}$,
M.\ Bonavita$^{20}$,
G.\ D'Ago$^{21}$
\newauthor
M.\ Dominik$^{17}$,
U.\ G.\ J{\o}rgensen$^{22}$, 
C.\ Snodgrass$^{20}$,
P.\ Spyratos$^{5}$,
L.\ Mancini$^{23,24,25}$
\\
$^{1}$\,Ankara University, Faculty of Science, Astronomy \& Space Sciences Department, Tandogan, TR-06100, Ankara, T\"urkiye\\
$^{2}$\,Ankara University, Astronomy and Space Sciences Research and Application Center (Kreiken Observatory),\\
{\.I}ncek Blvd., TR-06837, Ahlatlıbel, Ankara, T\"urkiye\\
$^{3}$\,Ankara University, Graduate School of Natural \& Applied Sciences, Astronomy \& Space Sciences Department,\\ Ziraat Mahallesi {\.I}rfan Ba\c{s}tu\u{g} Caddesi, D{\i}\c{s}kap{\i}, TR-06110 Alt{\i}nda\u{g} / Ankara, T\"urkiye\\
$^{4}$\,School of Mathematics, University of Leeds, Leeds LS2 9JT, UK \\
$^{5}$\,Astrophysics Group, Keele University, Staffordshire, ST5 5BG, UK\\
$^{6}$\,Astrobiology Research Unit, University of Li\`ege, All\'ee du 6 ao\^ut, 19, 4000 Li\`ege (Sart-Tilman), Belgium \\
$^{7}$\,Department of Earth, Atmospheric and Planetary Science, Massachusetts Institute of Technology, 77 Massachusetts Avenue, \\
Cambridge, MA 02139, USA\\
$^{8}$\,Instituto de Astrof\'isica de Canarias (IAC), Calle V\'ia L\'actea s/n, 38200, La Laguna, Tenerife, Spain \\
$^{9}$\,Baronnies Proven\c{c}ales Observatory, Hautes Alpes, Parc Naturel Regional des Baronnies Proven\c{c}ales, F-05150 Moydans, France\\
$^{10}$\,Universit{\"a}t Hamburg, Faculty of Mathematics, Informatics and Natural Sciences, Department of Earth Sciences, Meteorological \\ Institute, Bundesstra\ss{}e 55, 20146 Hamburg, Germany \\
$^{11}$\,School of Physics \& Astronomy, University of Birmingham, Edgbaston, Birmingham B15 2TT, UK \\
$^{12}$\,Space Sciences, Technologies and Astrophysics Research (STAR) Institute, Universit\'e de Li\`ege, All\'ee du 6 Ao\^ut 19C, B-4000\\ Li\`ege, Belgium \\
$^{13}$\,Instituto de Astronomia y Ciencias Planetarias, Universidad de Atacama, Copayapu 485, Copiapo, Chile \\
$^{14}$\,Millennium Institute of Astrophysics MAS, Nuncio Monsenor Sotero Sanz 100, Of.\ 104, Providencia, 7510109 Santiago, Chile \\
$^{15}$\,Instituto de Astrof\'{\i}sica, Facultad de F\'{\i}sica, Pontificia Universidad Cat\'olica de Chile, Av.\ Vicu\~na Mackenna 4860, 7820436, \\ Macul, Santiago, Chile \\
$^{16}$\,Centre for Exoplanet Science, SUPA, School of Physics \& Astronomy, University of St Andrews, North Haugh, St Andrews \\ KY16 9SS, UK \\
$^{17}$\,University of Southern Denmark, Department of Physics, Chemistry and Pharmacy, SDU-Galaxy, Campusvej 55, 5230, Odense M, \\ Denmark \\
$^{18}$\,Centro de Astronom{\'{\i}}a, Universidad de Antofagasta, Av.\ Angamos 601, 02800 Antofagasta, Chile \\
$^{19}$ \,Perimeter Institute for Theoretical Physics, 31 Caroline St N, Waterloo, ON N2L 2Y5, Canada \\
$^{20}$\,Institute for Astronomy, University of Edinburgh, Royal Observatory, Edinburgh EH9 3HJ, UK \\
$^{21}$\, Institute of Astronomy, University of Cambridge, Madingley Road, Cambridge CB3 0HA, United Kingdom\\
$^{22}$\,Centre for ExoLife Sciences, Niels Bohr Institute, University of Copenhagen, {\O}ster Voldgade 5, 1350 Copenhagen, Denmark \\
$^{23}$\,Department of Physics, University of Rome ``Tor Vergata'', Via della Ricerca Scientifica 1, 00133 Rome, Italy\\
$^{24}$\,INAF -- Turin Astrophysical Observatory, via Osservatorio 20, 10025 Pino Torinese, Italy\\
$^{25}$\,Max Planck Institute for Astronomy, K\"{o}nigstuhl 17, 69117 Heidelberg, Germany
}
\date{Accepted XXX. Received YYY; in original form ZZZ}

\pubyear{2025}

\begin{document}
\label{firstpage}
\pagerange{\pageref{firstpage}--\pageref{lastpage}}
\maketitle

\begin{abstract}
  WASP-4\,b is a hot Jupiter exhibiting a decreasing orbital period, prompting investigations into potential mechanisms driving its evolution. We analyzed 173 transit light curves, including 37 new observations, and derived mid-transit timings with {\sc exofast}, forming the most extensive TTV dataset for this system. Adding 58 literature timings and removing unreliable data, we constructed a TTV diagram with 216 points. Our analysis considered linear, quadratic, and apsidal motion models, with the quadratic model proving to be significantly superior in all model comparison statistics. We found no significant periodic signals in the data. The quadratic model allows us to infer a tidal quality factor of Q$^{\prime}_{\star} \sim 80,000$ from the orbital decay rate if this is due to stellar tides. Theoretical considerations indicate that such efficient dissipation is possible due to internal gravity waves in the radiative core of WASP-4, but only in our models with a more evolved host star, possibly near the end of its main-sequence lifetime, and with a larger radius than the observed one. Our main-sequence models produce only about a third of the required dissipation  (Q$^{\prime}_{\star}\sim 2 - 5 \times 10^5$). Therefore, the observed orbital decay can only be explained by a slightly larger or more evolved host, resembling the case for WASP-12. Our findings highlight the need for further stellar modeling and improvement in our current understanding of tidal dissipation mechanisms driving orbital decay in close-in exoplanetary systems.
\end{abstract}




\section{Introduction}
\label{sec:introduction}
WASP-4\,b was one of the first exoplanets discovered by the SuperWASP project \citep{wilson2008}. It is an inflated hot Jupiter ($R_{\rm p} = 1.321 \pm 0.039$\,R$_{\rm jup}$) on a short-period orbit ($P\approx 1.34$ d), making it a good candidate both for studying its tidal interactions with its host star and its atmosphere. That is why it has been studied extensively to date, and its transits have been observed with very high precision in different passbands for more than 15 years since its discovery. These long-term datasets enable the detection of subtle transit timing variations (TTVs), potentially indicative of orbital decay driven by stellar tides.

\citet{hoyer2013} constructed and analyzed the first TTV diagram of the target, based on 12 new transit observations and previous observations by \citet{wilson2008,gillon2009,winn2009,southworth2009,dragomir2011,sojeda2011,nikolov2012}, to improve constraints on the system's parameters. Their analysis did not reveal any TTVs with a root mean square (rms) amplitude exceeding 20 seconds over a four-year duration. 

\citet{bouma2019} performed a detailed analysis of this system based on the transit light curves of \citet{hoyer2013} and \citet{huitson2017} as well as occultation observations from Spitzer \citep{caceres2011,zhou2015}. They eliminated spot-crossing events, apsidal precession, and additional bodies in the system as potential causes of the observed TTVs. The orbital decay model turned out to be a slightly better fit to the data than apsidal precession, after ruling out potential systematic errors. From their quadratic model, they found the rate of the secular period change $\dot{\rm P} = -(12.6 \pm 1.2)$ ms yr$^{-1}$, and the modified tidal quality factor of the star $Q^{\prime}_{\star} = (2.9 \pm 0.3) \times 10^4$ based on their orbital decay model. 

\cite{southworth2019} added 22 new transit light curves to the TTV diagram that they updated and analyzed, finding a period change of $\dot{P} = -9.2 \pm 1.1$~ms~yr$^{-1}$. They ruled out the Applegate mechanism \citep{applegate1992}, as \citet{bouma2019} did previously, because it would only lead to a TTV shift of 15 seconds over a 50 year-cycle \citep[as suggested by][]{watsonmarsh2010}, which is significantly smaller than the observed shift in transit timings. They also ruled out an outside perturber as the cause, because even when the observed TTV is assumed to be due to a longer period perturber, they deduced from its amplitude that it should have been observed in the radial velocities (hereafter RVs). Although they did not reject orbital decay as a potential cause, they did not find any evidence for it in their analysis and provided Q$^{\prime}_{\star} > 10^{4.58}$ as a lower limit. The measured $\dot{P}$ was found to correspond to a tidal quality factor of $Q^{\prime}_{\star} = (3.8 \pm 0.3) \times 10^4$.

\citet{baluev2019} homogeneously analyzed the transit light curves of WASP-4\,b and created a TTV diagram based on their analysis of mid-transit timings, but did not find a departure from a linear ephemeris. They also analyzed the RVs for WASP-4, and found the RV trend estimation consistent with zero or as small as -4 m s$^{-1}$ yr$^{-1}$ in their 3-$\sigma$ limits. This would indicate a TTV trend due to orbital decay with a timescale of 70 Myrs, which is consistent with the observed transit times within 2$\sigma$. Therefore, they cannot conclusively rule out or confirm such a trend in their data.

Following \citet{bouma2019}, \citet{bouma2020} combined their new RV observations with previous measurements and found a significant line-of-sight acceleration of $\dot{v_{\gamma}} = -0.0422 \pm 0.0028$ m s$^{-1}$ day$^{-1}$, which translates into an expected period decrease of -5.9 ms yr$^{-1}$, which is comparable to the period decrease measured from transit timings based on their dataset in 2019 \citep{bouma2019}. They calculated a minimum mass for the object that would induce the linear trend in RVs following \citet{feng2015}, and their results showed that the companion responsible for such an acceleration should have a true mass of 10–300 M$_{\rm jup}$ and an orbital distance of 10–100 au.

\citet{baluev2020} re-analyzed the RV observations of the system, also including CORALIE and HARPS datasets not analyzed by \citet{bouma2020}. Since it was only the Keck observations from \citet{bouma2020} that deviated from the constant systemic velocity ($v_{\gamma}$), they rejected the acceleration model of \citet{bouma2020}. They considered only orbital decay as a potential explanation, considering apsidal precession to be unlikely for this particular system given its eccentricity, which is consistent with zero based on host star RVs. Based on the orbital decay scenario, they found $Q^{\prime}_{\star} \sim 60 000$ within a 1$\sigma$ range of 45,000–85,000.

\citet{maciejewski2022} published a TTV analysis of WASP-4\,b, in which they found that a quadratic trend in the data indicating a decrease in the orbital period by a rate of dP$_{\rm orb}$ / dE = $(-2.1 \pm 0.6) \times 10^{-10}$ days per orbit is superior to their linear model with a $\Delta$ BIC value of 6.7, as a result of which they encouraged future observations of the system.

\citet{turner2022} added sector-28 and 29 observations from TESS to the TTV dataset and analyzed these together with a homogeneously measured set of RVs with a non-standard pipeline. Their RV model based on a circular orbit showed no evidence for an acceleration of the centre of mass of the system with respect to the observer. Finally, their two-planet solution of the RV data ended up with better fit statistics, with a secondary of mass $M_c = 5.47$ M$_{\rm jup}$ at 6.82 au. This companion is unlikely to cause a potential von Zeipel-Kozai-Lidov cycle, since the lower mass limit for this is only satisfied when its orbital inclination is below 2.5$^\circ$ \citep[as calculated by][]{bouma2019}. It cannot be the source of the observed TTVs either because it cannot induce an amplitude larger than $\sim2$ seconds in total based on its parameters \citep[as found by][from their RV model]{turner2022} to be M$_c = 5.47$ M$_{\rm jup}$ and $P_c = 7001.0$ days. As a result, they found that orbital decay is the best model in explaining the TTV data with a decay-rate of $\dot{P} = -7.33 \pm 0.71$ ms yr$^{-1}$ and a decay-timescale $\tau = 15.77$ Myrs. From the best fitting quadratic coefficient, they found the modified tidal quality factor $Q^{\prime}_{\star} = (5.1 \pm 0.9) \times 10^4$, which probably either implies a host-star that is leaving the main-sequence or that an exterior planet is trapping WASP-4\,b's spin vector in a high-obliquity state. However, since it has been found to be in a low-obliquity configuration, WASP-4\,c is unlikely the cause even if it exists. 

\citet{harre2023a} added eight light curves from the CHaracterising ExOPlanet Satellite (CHEOPS) to the dataset and re-analyzed TTV data for WASP-4\,b, as a result of which they found the orbital decay model is statistically superior to their apsidal precession and Keplerian orbit models. Occultation measurements, on the other hand, were inconclusive in terms of the apsidal precession model due to the scatter of the data and their rather large error bars. They found an orbital decay rate of d$P$/d$E$ = $(-2.62 \pm 0.49) \times 10^{-10}$ days per orbit, which corresponds to a modified tidal quality factor $Q^{\prime}_{\star} = (5.7 \pm 1.0) \times 10^4$, and a decay timescale of $\tau = 18.73 \pm 3.48$ Myrs.

\citet{harre2023b} extended the dataset from \citet{harre2023a} with sector-49 observations of TESS. They investigated if the candidate planetary companion on a 7001 day-orbit, as established by \citet{turner2022}, would induce observable orbital motion of WASP-4. To include a correction of the light-time effect (LiTE), they applied the formula of \citet{schneider2005} to find the maximum time-shift. Using the candidate's parameters suggested by \citet{turner2022}, they found a maximum shift $\Delta T_{\rm max} = 37.7$ seconds to validate it and obtained a 37.6 second shift with a dynamical approach, assuming a circular orbit. After they corrected TTVs for the LiTE due to \citet{turner2022}'s planet-c, assuming three different values for its conjunction time ($T_{O,c}$), they found very similar fit statistics with and without LiTE-correction due to planet-c for orbital decay and apsidal precession models. However, they did not find a conclusive answer to the question of what is the true origin of the TTVs of WASP-4 b. Most recently, \citet{Ma_hst2025} added the mid-transit time they obtained from the transit light curve acquired with the Hubble Space Telescope (HST). At the end of their analysis of the TTV diagram, composed of literature, TESS and HST transit timings, the orbital decay model proved to be statistically superior with a rate of $\dot{P} = -6.46 \pm 0.58$ ms yr$^{-1}$, giving an estimate for the tidal quality parameter $Q^{\prime}_{\star} > 5.7 \times 10^4$.

We investigated the orbital decay potential of WASP-4\,b through the analysis of its TTV diagram based on a very large dataset of mid-transit timings, measured from the light curves appeared in the literature and open databases for ground-based amateur and space-borne data as well as from our own observations combined with the mid-transit timings reported in the literature, that are derived from light curves we did not have access to. We present detailed information of observations and data reduction in Section \ref{sec:observations}, our analyses for a global model (Sec \ref{subsec:global_modelling}) to derive system parameters and TTV models (Sec \ref{subsec:ttv_analysis}) to determine transit timing behaviour. We discuss the implications of our TTV analysis in terms of tidal interactions between the host star and the close-in gas giant in Section \ref{sec:discussion} and finally provide a summary of the work and our conclusions in Section \ref{sec:conclusion}.
 
\section{Observations and Data Reduction}
\label{sec:observations}

\subsection{Observations with the Danish Telescope in La Silla}
\label{subsec:lasilla}

\begin{table*} \centering
\caption{\label{tab:obslog} Log of the transit observations obtained for WASP-4 from the Danish telescope, TRAPPIST-South Telescope, and El Sauce Obs. CDK20 Telescope. $N_{\rm obs}$ is the number of observations, $T_{\rm exp}$ is the exposure time, $T_{\rm dead}$ is the mean time between the end of one exposure and the start of the next, `Moon illum.' is the fractional illumination of the Moon at the midpoint of the transit, and $N_{\rm poly}$ is the order of 
the polynomial fitted to the out-of-transit data. The aperture radii are target aperture, inner sky and outer sky, respectively.}
\setlength{\tabcolsep}{4pt}
\begin{tabular}{lccccccccccccc} \hline
Date of first & Start time & End time  &$N_{\rm obs}$ & $T_{\rm exp}$ & $T_{\rm dead}$ & Filter & Airmass &  Moon & Aperture & $N_{\rm poly}$ & Scatter \\
observation   &    (UT)    &   (UT)    &              &      (s)      &      (s)       &        &  & illumination & (pixels) &                & ($m$mag)  \\
\hline
\multicolumn{4}{l}{Observations with the 1.54\,m Danish Telescope} \\
\hline
2019/07/17 & 03:16 & 07:38 & 136 & 100      & 13 & $R$ & 2.12 $\to$ 1.02 $\to$ 1.05 & 0.999 & 10 30 40 & 2 & 1.86 \\
2019/07/25 & 04:00 & 07:57 & 126 & 100      & 13 & $R$ & 1.56 $\to$ 1.02            & 0.483 & 16 24 50 & 2 & 1.10 \\
2019/07/29 & 02:25 & 07:13 & 101 & 100      & 14 & $R$ & 1.47 $\to$ 1.02            & 0.112 & 16 25 50 & 1 & 1.12 \\
2019/08/02 & 04:46 & 08:44 & 124 & 100      & 14 & $R$ & 1.23 $\to$ 1.02 $\to$ 1.25 & 0.020 & 17 27 50 & 2 & 1.27 \\
2019/08/06 & 04:41 & 08:57 & 134 & 100      & 14 & $R$ & 1.21 $\to$ 1.02 $\to$ 1.23 & 0.342 & 11 19 40 & 1 & 0.97 \\
2019/08/10 & 04:58 & 09:14 & 133 & 100      & 14 & $R$ & 1.12 $\to$ 1.02 $\to$ 1.15 & 0.760 & 12 20 50 & 2 & 0.84 \\
2019/08/14 & 05:48 & 09:45 & 124 & 100--90  & 14 & $R$ & 1.05 $\to$ 1.02 $\to$ 1.27 & 0.986 & 12 20 50 & 1 & 1.59 \\
2019/08/18 & 05:46 & 09:59 & 134 & 100      & 13 & $R$ & 1.04 $\to$ 1.02 $\to$ 1.38 & 0.929 & 12 20 40 & 2 & 1.07 \\
2019/08/22 & 05:35 & 10:21 & 136 & 100      & 14 & $R$ & 1.03 $\to$ 1.02 $\to$ 1.57 & 0.629 & 15 24 50 & 2 & 1.09 \\
2019/08/26 & 06:31 & 09:47 & 103 & 100      & 14 & $R$ & 1.03 $\to$ 1.67            & 0.223 & 15 25 50 & 1 & 0.94 \\
2019/08/30 & 06:01 & 10:00 & 115 & 100      & 13 & $R$ & 1.03 $\to$ 1.02 $\to$ 1.76 & 0.001 & 17 26 50 & 1 & 0.77 \\
2021/08/07 & 05:17 & 08:21 & 101 & 100--60  & 16 & $R$ & 1.11 $\to$ 1.02 $\to$ 1.11 & 0.019 & 16 25 50 & 1 & 0.78 \\
2021/09/15 & 00:22 & 04:32 & 132 & 100      & 13 & $R$ & 1.64 $\to$ 1.02            & 0.641 & 14 24 50 & 2 & 1.38 \\
2021/09/19 & 00:49 & 04:27 & 107 & 100      & 14 & $R$ & 1.40 $\to$ 1.02            & 0.959 & 11 20 40 & 1 & 4.69 \\
2021/09/23 & 01:35 & 05:30 & 116 & 100      & 14 & $R$ & 1.17 $\to$ 1.02 $\to$ 1.18 & 0.953 & 13 22 50 & 2 & 1.27 \\
2021/09/27 & 01:48 & 05:19 & 110 & 100      & 15 & $R$ & 1.13 $\to$ 1.02 $\to$ 1.08 & 0.680 & 13 23 50 & 3 & 0.78 \\
2021/10/05 & 01:54 & 06:14 & 164 & 100--80  & 15 & $R$ & 1.07 $\to$ 1.02 $\to$ 1.25 & 0.024 & 15 35 50 & 2 & 1.43 \\
2021/10/13 & 01:45 & 06:59 & 161 & 100      & 14 & $R$ & 1.05 $\to$ 1.02 $\to$ 1.61 & 0.509 & 17 27 50 & 2 & 1.29 \\
2021/10/17 & 03:25 & 07:12 & 107 & 100      & 28 & $R$ & 1.04 $\to$ 1.83            & 0.885 & 13 23 45 & 1 & 0.90 \\
2022/09/22 & 01:57 & 05:00 & 107 & 100--80  & 14 & $R$ & 1.15 $\to$ 1.02 $\to$ 1.04 & 0.141 & 17 25 50 & 2 & 0.94 \\
2022/10/12 & 03:02 & 07:02 & 118 & 100--120 & 14 & $R$ & 1.02 $\to$ 1.60            & 0.940 & 12 21 40 & 2 & 1.25 \\
2023/08/05 & 04:26 & 09:00 & 142 & 100      & 16 & $R$ & 1.26 $\to$ 1.02 $\to$ 1.09 & 0.825 & 18 27 60 & 2 & 0.81 \\
2023/08/13 & 05:41 & 09:44 & 124 & 100      & 15 & $R$ & 1.06 $\to$ 1.02 $\to$ 1.26 & 0.086 & 18 27 60 & 1 & 0.58 \\
2023/08/17 & 06:09 & 09:32 & 108 & 100      & 15 & $R$ & 1.03 $\to$ 1.02 $\to$ 1.37 & 0.009 & 17 27 60 & 1 & 0.89 \\
2023/09/29 & 02:03 & 05:41 & 113 & 100      & 14 & $R$ & 1.10 $\to$ 1.02 $\to$ 1.14 & 0.999 & 21 30 60 & 2 & 0.75 \\
2024/07/26 & 04:18 & 07:36 & 101 & 100      & 17 & $R$ & 1.43 $\to$ 1.03            & 0.708 & 18 28 60 & 1 & 0.69 \\
2024/07/30 & 02:58 & 08:45 & 180 & 100--90  & 14 & $R$ & 1.87 $\to$ 1.02 $\to$ 1.05 & 0.270 & 16 26 50 & 2 & 1.03 \\
2024/09/27 & 01:45 & 05:43 & 112 & 100--80  & 14 & $R$ & 1.14 $\to$ 1.02 $\to$ 1.11 & 0.267 & 18 28 50 & 1 & 1.07 \\
\hline
\multicolumn{4}{l}{Observations from El Sauce Observatory} \\
\hline
2024/07/14 & 04:04 & 07:29 & 84 & 120 & 20 & L & 1.78 $\to$ 1.05 & 0.53 & 14 20 25 & - & 2.36 \\
2024/07/18 & 03:11 & 07:50 & 115 & 120 & 20 & L & 2.16 $\to$ 1.02 & 0.88 & 14 20 25 & - & 2.22 \\
2024/08/27 & 06:43 & 10:14 & 81 & 120 & 20 & R & 1.04 $\to$ 1.69 & 0.88 & 12 20 25 & - & 4.29 \\
2024/09/03 & 00:43 & 03:55 & 60 & 120 & 20 & R & 1.77 $\to$ 1.07 & 0.05 & 12 20 25 & - & - \\
\hline
\multicolumn{4}{l}{TRAPPIST-South Observations} \\
\hline
2024/08/23 & 06:02 & 10:24 & 309 & 40 & 11 & $z'$ & 1.03 $\to$ 1.63 & 0.91 & 7 20 32 & - & 2.49 \\
2024/08/27 & 06:00 & 10:24 & 313 & 40 & 11 & $z'$ & 1.03 $\to$ 1.74 & 0.41 & 7 20 32 & - & 2.54 \\
\hline
\multicolumn{4}{l}{Ckoirama Observations} \\
\hline
2019/09/25 & 01:31 & 06:44 & 95  & 140-150 & 2 & $r'$ & 1.21 $\to$ 1.05 $\to$ 1.16 & 0.103 & 8 18 22 & 1 & 2.25 \\
2019/09/29 & 02:00 & 06:02 & 75  & 180 & 2 & $r'$ & 1.28 $\to$ 1.05 $\to$ 1.22 & 0.031 & 13 18 28 & 1 & 1.43 \\
2019/10/03 & 01:56 & 06:37 & 87  & 180 & 2 & $r'$ & 1.28 $\to$ 1.05 $\to$ 1.39 & 0.347 & 8 14 20  & 1 & 1.25 \\
2019/10/07 & 01:52 & 07:01 & 95  & 180 & 2 & $r'$ & 1.09 $\to$ 1.05 $\to$ 1.60 & 0.734 & 9 14 19  & 1 & 1.44 \\
2019/10/11 & 01:26 & 07:00 & 104 & 180 & 2 & $r'$ & 1.11 $\to$ 1.05 $\to$ 1.71 & 0.972 & 8 13 18  & 1 & 0.99 \\
\hline \end{tabular}
\end{table*}

We observed 28 transits of WASP-4 in the 2019 and 2021--24 observing seasons using the 1.54\,m Danish Telescope at ESO La Silla, Chile (Table\,\ref{tab:obslog}). No observations were obtained in 2020 due to the Covid-19 pandemic. Our observing approach was the same as that used in \citet{southworth2019}, in order to maximise the homogeneity of the data. In brief, we used the Danish Faint Object Spectrograph and Camera (DFOSC) in imaging mode, defocused the telescope to increase the precision of the data \citep{southworth2009_wasp5}, windowed the CCD to decrease the readout time, but did not apply pixel binning. Exposure times of 100\,s were used, shortened in some cases to avoid image saturation, through a Bessell $R$ filter for high throughput.

The data were reduced using the {\sc defot} pipeline \citep{southworth2014}, which uses the {\sc idl}\footnote{{\tt https://www.ittvis.com/idl/}} implementation of the {\sc aper} routine from {\sc daophot} \citep{Stetson87pasp} contained in the NASA {\sc astrolib}\footnote{{\tt https://asd.gsfc.nasa.gov/archive/idlastro/}} library to perform aperture photometry. Bias and flat-field calibrations were constructed but not used because their main effect was to increase the scatter of the resulting light curves. A differential-magnitude light curve was generated for each transit observation by subtracting the magnitude of a composite comparison star from that of the target star. A polynomial was fitted to the out-of-transit data, with its coefficients optimised simultaneously with the weights of the comparison stars used. The polynomial order was either 1 or 2 (Table\,\ref{tab:obslog}), with a single instance of order 3 for a night where the sky conditions changed a lot during the observations. We discarded the light curve we acquired on 19 September 2021 because it is incomplete due to bad weather conditions at the time.

Timing checks are frequently performed at the DK 1.54m telescope as part of the workflow when observing exoplanetary transits. The timestamps for the midpoint of each image were taken from the headers of the {\sc fits} files and converted to the BJD reference frame with the dynamical (TDB) time standard component as BJD$_{\rm TDB}$ using routines from \citet{eastman2010}. Manual time checks were performed for many transits, in all cases confirming the correctness of the timestamps to within typically 1\,s.

\subsection{Transit Observations at the Ckoirama Observatory}

We obtained five transit light curves of WASP-4\,b with the 0.6\,m Chakana telescope, situated at the Ckoirama Observatory \citep{char2016} in northern Chile—the inaugural state-operated astronomical facility in the region—is under the management and operational oversight of the Centro de Astronomía (CITEVA) at the Universidad de Antofagasta. This f/6.5 telescope is equipped with an FLI Pro-Line 16801 CCD camera, utilised here with a selection of Sloan photometric filters (u$^{\prime}$, g$^{\prime}$, r$^{\prime}$, and i$^{\prime}$). In its unbinned configuration, this setup yields a field of view of $32^{\prime}.4\times32^{\prime}.4$ with a corresponding pixel scale of 0.47 arcsec pixel$^{-1}$. The photometric data that were acquired at Ckoirama used the Sloan r$^{\prime}$ filter, employing 2$\times$2 binning to enhance the signal-to-noise ratio for comparison stars, and reduce the CCD readout to 1.8\,s. A detailed log of the Chakana observations is provided in Table \ref{tab:obslog}. The data were reduced using the same methodology as the Danish data, in that aperture photometry was performed using the defot pipeline \citep{southworth2014} and bias and flat field calibrations were performed. In all cases, a first order polynomial was used to detrend the resulting differential magnitude light curve. The midpoint timestamps of each image were then converted to BJD TDB using routines from \citet{eastman2010}.

\subsection{TRAPPIST-South Observations}
\label{subsec:trappist}
We observed two full transits with the TRAPPIST-South \citep[TRAnsiting Planets and PlanetesImals Small Telescope,][]{Jehin2011,Gillon2011} telescope on UTC August 23 and 27, 2024 in the Sloan-$z'$ filter with an exposure time of 40s. The TRAPPIST-South is a 60-cm Ritchey-Chretien telescope located at ESO-La Silla Observatory in Chile. It is equipped with a thermoelectrically cooled 2K$\times$2K FLI Proline CCD camera with a field of view of $22\arcmin\times22\arcmin$ and a pixel-scale of 0.65\arcsec/pixel. Data reduction and photometric extraction were performed using the {\sc prose} pipeline \citep{prose}. All timings were converted to BJD$_{\rm TDB}$.

\subsection{Transit Observations at the El Sauce Observatory}
\label{subsec:el_sauce}
Four transits were observed in 2024 on July 14 and 18, August 27, and September 3 using a 0.5m CDK20 telescope located at El Sauce Observatory (IAU observatory code X02) in Chile but controlled remotely from the Observatory of Baronnies Provençales (OBP, IAU observatory code B10) in France. The OBP is a private observatory doing outreach, courses, training, and research, addressed to all public, amateur, and professional astronomers. The CDK20 telescope is on a paramount equatorial mount and is equipped with a Moravian G4 16K CCD camera. The photometric data were obtained from 120s exposure frames, after standard calibration. A Luminance filter was used to maximize the SNR. The images were taken with $1\times1$ binning and photometry was performed with an aperture of 9 pixels, with the FWHM of the target estimated to $2.3^{{\prime}{\prime}}$ with a pixel scale of $0.5255^{\prime\prime}/\textnormal{pixel}$. The analysis was performed using the {\sc Muniwin} program from the photometry software package {\sc C-Munipack} \citep{hroch2014}. Observation times were recorded in JD$_{\rm UTC}$ and then converted to BJD$_{\rm TDB}$ based on the observatory location and the equatorial coordinates of the target. Both observations benefited from good weather conditions, but the low elevation of the target did not permit us to observe the plateau expected before transit on July 14th. The lower SNR we observed on July 18th is due to light pollution from the moon. We had to eliminate the light curve we acquired on September 3rd because we missed the ingress and had to start the observations around the second contact, the timing of which we failed to constrain.

\subsection{TESS Observations}
\label{subsec:tess}
TESS observed WASP-4 during sectors 2, 28, 29, and 69; between September 2018 and September 2023 in 2-minute cadence mode. We obtained products of the Science Processing Operations Center (SPOC) pipeline \citep{jenkins2016} from the Barbara A. Mikulski Archive for Space Telescopes (MAST) of the Space Telescope Science Institute (STScI) of NASA. We examined the Simple Aperture Photometry (SAP), Presearch Data Conditioning SAP (PDCSAP), and Data Validation Timeseries (DVT) light curve products with the {\sc lightkurve} Python package \citep{lightkurve2018}. We then examined target pixel files (TPF) to check for contamination in the apertures SPOC made use of and experimented with different apertures for both the target and the background. We haven't detected any extra light other than the target, Gaia  lists no source in a 30$^{\prime\prime}$ radius \citep{gaia2016,gaia2023a}, and the TIC contamination ratio is only 1.1\%. We also have not achieved better light curves in our experiments with different apertures than the SPOC's DVT, which has proven their robust use to derive mid-transit times and system parameters \citep{riddenharper2020,basturk2022,yalcinkaya2024}, because they are detrended from the correlated noise sources. 


\subsection{Summary of Observations and Literature Data}
\label{subsec:otherobs}
We also collected all the high-quality, full transit light curves from the Exoplanet Transit Database (ETD) (41 in total) and the literature (34 in total), spanning a time range of 15.98 years. Together with 61 DVT light curves from TESS, we analyzed a total of 136 light curves, which have been visually checked by us for their quality before they were used for the analysis. We also eliminated incomplete light curves from our sample. We corrected the light curves for the undetrended, linear airmass effects only where it was necessary by making use of the software package AstroImageJ (AIJ) \citep{collins2017}. All the observation times have been converted to BJD$_{\rm TDB}$ by making use of the relevant {\sc astropy} constants and functions \citep{astropy2013,astropy2018} when they are provided in other timing reference frames.    

We further selected transit light curves based on the photon-noise rate (PNR) \citep{fulton2011} and $\beta$-factor statistics \citep{winn2008}, which are frequently used to characterize white and red noise, respectively. We removed the light curves that have PNR values higher than the transit depth, while we eliminated those with $\beta$-factors larger than 2.5, as we did in \citet{basturk2022} and \citet{yalcinkaya2024}. Light curve asymmetries caused by spot-crossing events are apparent in several cases. Heavily affected light curves — such as those obtained with DK154 on 2021-09-23 and 2021-10-05 (Fig. \ref{fig:transits_02})  — were excluded from the analysis due to their large $\beta$-factors. However, some light curves exhibiting less prominent spot-induced anomalies remained in the sample, as they did not exceed the exclusion threshold. Light curves that were eliminated from further analysis, and the reasons for their elimination, are also provided in these tables. During this elimination procedure, five light curves from ETD, seven from the literature, and three that we acquired in our own observations were eliminated, leaving us with a total of 158 light curves. We provide the entire dataset in five separate tables as online material, first few lines of which are provided in Appendix \ref{app:midtransits}. 

\begin{figure}
    \centering
    \includegraphics[width=\columnwidth]{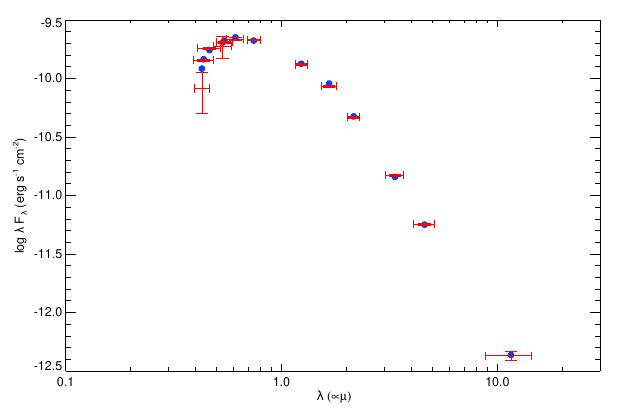}
    \caption{Broadband fluxes of WASP-4. Red points represents the data points including error bars, while blue points represents model fluxes. The error bars in wavelength denote the bandwidth of the corresponding filter.}
    \label{fig:sedplot}
\end{figure}

\begin{figure*}
\includegraphics[width=\textwidth]{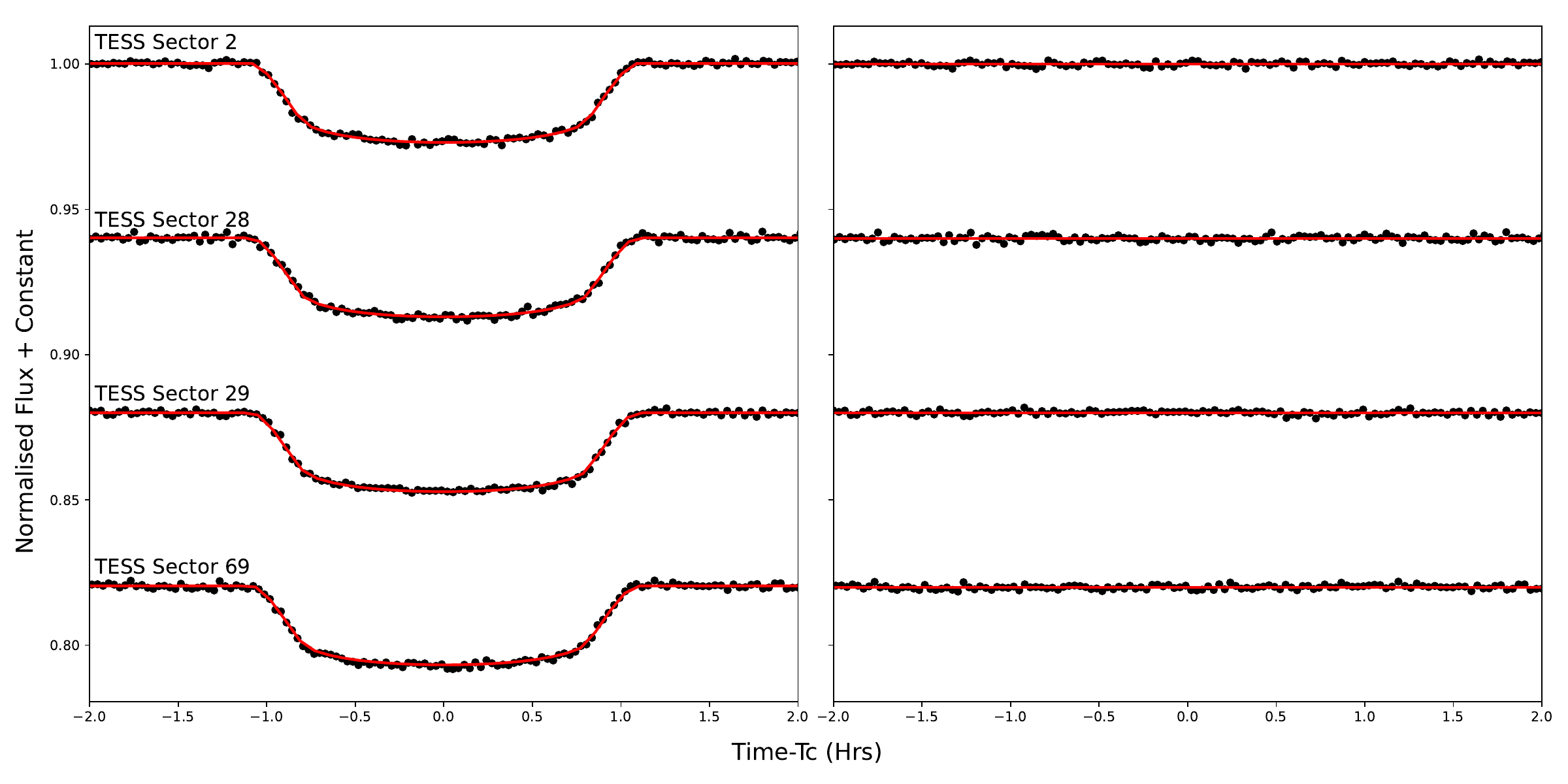}
    \caption{Left panel: TESS transit light curves of WASP-4 b used in global modelling (black dots) and the {\sc exofastv2} model (red continuous line). Each TESS sector is time-folded and binned in two-minute intervals. Right panel: residuals from the global model (black dots). The red line indicates the zero level.}
    \label{fig:gm_lcs}
\end{figure*}

\section{Analysis and Results}
\label{sec:analysis}
\subsection{Global Modelling}
\label{subsec:global_modelling}
To explore the potential consistency or otherwise between the observed (modified) tidal quality factor ($Q^{\prime}$) and theoretical predictions, we need to constrain the stellar age, metallicity and stellar mass ($M_{\star}$) \citep{2024MNRAS.527.5131B}. Metallicity can be calculated from high-resolution spectral analysis, but determining the age for cold stars like WASP-4 is challenging because their mass and radius change only very slightly during their main sequence (MS) evolution. In fact, for cold stars, the mass is considered almost constant ($\dot{M} \sim 10^{-4}$ M$_{\odot}$ Gyr$^{-1}$ for a solar metallicity 1 M$_{\odot}$ star \citep{choi2016} during its MS evolution, making the radius a key parameter for accurate determination of the age. The stellar radius, on the other hand, can be accurately calculated by performing a spectral energy distribution (SED) analysis based on a precise parallax value from \textit{Gaia} \citep{stassun2018}. Hence, we first modelled the SED of WASP-4 in order to acquire the stellar radius and also effective temperature ($T_{\rm eff}$) with the {\sc exofastv2} \citep{eastman2017, eastman2019} code and the Modules for Experiments in Stellar Astrophysics ({\sc MESA}) Isochrones \& Stellar Tracks (MIST) bolometric correction grid \citep{choi2016}. For this model, we used the atmospheric parameters of WASP-4 from \cite{wilson2008} as Gaussian priors and modelled the broadband magnitudes of the star given in Table \ref{tab:sed}. We supplied the \textit{Gaia} DR3 parallax \citep{gaia2016,gaia2021a,gaia2021b} in order to calculate the stellar luminosity after correcting for the parallax offset \citep{lindegren2021}. During SED fitting, interstellar extinction ($A{_{V}}$) along the line of sight was limited to the value given by \citet{schlegel1998}. From this analysis, we obtain $R_{\star, \rm SED}= 0.9150^{+0.0089}_{-0.0091}$ R$_{\odot}$ and $T_{\rm eff,SED}=5488^{+29}_{-28}$ K. The SED model for broadband magnitudes is shown in Fig. \ref{fig:sedplot}. 

After acquiring the stellar radius, we fitted the {\sc MESA} isochrones with {\sc exofastv2} to determine the age and mass of the host star. This fit requires atmospheric parameters of the host star ($T_{\rm eff}$, $[{\rm Fe/H}]$ and $\log{g}$ or mean stellar density $\rho_{\star}$), which were reported by \citet{mortier2013}, \citet{doyle2013}, \citet{sousa2021}, \citet{petrucci2013}, \citet{gillon2009} and \citet{wilson2008} based on their spectral analyses. Although the reported values of these parameters are slightly different from each other, the resulting ages could differ by the evolutionary stage (MS or subgiant) the star is passing through. Therefore, we performed six different isochrone fits for each of the reported $T_{\rm eff}$, $[{\rm Fe/H}]$ pairs as Gaussian priors given in Table \ref{tab:globalmodel}. In these models, we used the $R_{\star, \rm SED}$ value from our SED analysis as a Gaussian prior, but instead of using a prior width from the SED analysis, we enforced 4\% of the median value as suggested by \citet{tayar2022}. Due to the subtle effect of surface gravity ($\log{g}$) on spectral lines (e.g. \citealt{2011ApJ...726...52H}), we did not use its value as a prior. Instead, we modelled the TESS light curves from DVT files together with MIST, which gives an accurate mean density value for the host star ($\rho_{\star}$) by making use of Kepler's third law \citep{2003ApJ...585.1038S, eastman2019}. The radial velocity semi-amplitude was measured by \cite{2017A&A...602A.107B} using all available RV measurements from different datasets. They also found that the planet's orbit is circular, so we fitted a circular model. We used the measured semi-amplitude as a Gaussian prior for calculating the planet's mass in each model. The fitting procedure stops when the parameters are considered ``well-mixed'', controlled by default values \citep{2006ApJ...642..505F} of the number of independent draws, $T_Z$ > 1000, which is the ratio of the length of the chains and their correlation length, and the Gelman-Rubin statistic $R_Z$ < 1.01, that quantifies the degree of similarity between independent chains. The posteriors of stellar parameters from six global models are listed in Table \ref{tab:globalmodel}. The TESS light curves with the global model are shown in Fig. \ref{fig:gm_lcs} while posteriors of transit parameters are listed in Table \ref{tab:transit params}. Although the results from different models agree within 1-$\sigma$, we favour Model-4 because the $T_{\rm eff}$ and $[{\rm Fe/H}]$ measurements were calculated from the spectra with the highest resolution and SNR among others, and the line list was meticulously selected as denoted by \cite{doyle2013}.

\begin{table}
\centering
\caption{Broadband apparent magnitudes for WASP-4.}
\label{tab:sed}
\begin{tabular}{ccc}  
\hline
\hline
Passband & $\lambda_{\rm eff}$ (nm) & Magnitude \\
\hline
\multicolumn{3}{l}{APASS-DR9 \citep{henden2016}}\\
\hline
Johnson $B$ & 437.81 & $13.216 \pm 0.020$\\
Johnson $V$ & 544.58 & $12.468 \pm 0.025$\\
SDSS $g'$ & 464.04 & $12.782 \pm 0.023$ \\
SDSS $r'$ & 612.23 & $12.283 \pm 0.033$ \\
SDSS $i'$ & 743.95 & $12.089 \pm 0.012$ \\
\hline
\multicolumn{3}{l}{2MASS \citep{cutri2003}}\\
\hline
$J$ & 1235 & $11.179 \pm 0.025$ \\ 
$H$ & 1662 & $10.842 \pm 0.026$ \\ 
$K_{S}$ & 2159 & $10.746 \pm 0.021$ \\ 
\hline
\multicolumn{3}{l}{All WISE \citep{cutri2013}}\\
\hline
$W_1$ & 3352.6 & $10.663 \pm 0.022$ \\ 
$W_2$ & 4602.8 & $10.728 \pm 0.020$ \\ 
$W_3$ & 11560.8 & $10.775 \pm 0.080$ \\
\hline
\multicolumn{3}{l}{Tycho-2 \citep{hog2000}}\\
\hline
$B_T$ & 428.0 & $13.792 \pm 0.414$ \\
$V_T$ & 534.0 & $12.597 \pm 0.231$ \\
\hline
\end{tabular}
\end{table}

\begin{table*}
\centering
\caption{Median values and 68\% confidence intervals for WASP-4 system and priors from our analysis.}
    \begin{adjustbox}{width=\textwidth}
        \begin{tabular}{llcccccc}
 \hline \hline
 Symbol & Parameter (Unit) & & & Values &  \\
\hline
\multicolumn{2}{l}{Priors:} & Model-1 & Model-2 & Model-3 & Model-4 & Model-5 & Model-6\\
\hline
$T_{\rm eff}$&Effective Temperature (K)&$5500\pm150$&$5500\pm100$&$5513\pm43$&$5400\pm90$&$5436\pm34$&$5496\pm19$\\
$[{\rm Fe/H}]$&Metallicity (dex) 
&$0\pm0.2$&$-0.03\pm0.09$&$0.03\pm0.03$&$-0.07\pm0.19$&$-0.05\pm0.04$&$0.05\pm0.01$\\
Reference & &\citet{wilson2008}&\citet{gillon2009}&\citet{mortier2013}&\citet{doyle2013}&\citet{petrucci2013}&\citet{sousa2021}\\
$R_{\star}$&Radius (R$_{\odot}$)&$0.915\pm0.037$&$0.915\pm0.037$&$0.915\pm0.037$&$0.915\pm0.037$&$0.915\pm0.037$&$0.915\pm0.037$\\
 \hline
 \multicolumn{2}{l}{Stellar parameters:}\\
 \hline
$M_{\star}$&Mass (M$_{\odot}$)&$0.882^{+0.06}_{-0.049}$&$0.878^{+0.051}_{-0.041}$&$0.895^{+0.046}_{-0.041}$&$0.858^{+0.053}_{-0.040}$&$0.862^{+0.043}_{-0.032}$&$0.899^{+0.044}_{-0.039}$\\
$R_{\star}$&Radius (R$_{\odot}$)&$0.903^{+0.02}_{-0.018}$&$0.902^{+0.017}_{-0.015}$&$0.907^{+0.016}_{-0.015}$&$0.895^{+0.017}_{-0.015}$&$0.896^{+0.015}_{-0.013}$&$0.908^{+0.015}_{-0.014}$\\
$L_{\star}$&Luminosity (L$_{\odot}$)&$0.671^{+0.069}_{-0.062}$&$0.669^{+0.055}_{-0.048}$&$0.684^{+0.034}_{-0.031}$&$0.624^{+0.045}_{-0.040}$&$0.633^{+0.027}_{-0.024}$&$0.677^{+0.025}_{-0.022}$\\
$\rho_{\star}$&Density (cgs)&$1.696^{+0.033}_{-0.04}$&$1.695^{+0.033}_{-0.040}$&$1.696^{+0.033}_{-0.041}$&$1.695^{+0.033}_{-0.037}$&$1.695^{+0.033}_{-0.039}$&$1.699^{+0.031}_{-0.040}$\\
$\log{g}$&Surface gravity (cgs)&$4.473^{+0.012}_{-0.011}$&$4.473\pm0.010$&$4.475^{+0.009}_{-0.010}$&$4.469^{+0.011}_{-0.010}$&$4.470^{+0.010}_{-0.09}$&$4.476^{+0.009}_{-0.010}$\\
$T_{\rm eff}$&Effective Temperature (K)&$5500^{+110}_{-110}$&$5496^{+81}_{-80}$&$5510^{+40}_{-39}$&$5422^{+71}_{-70}$&$5437^{+31}_{-31}$&$5496^{+17}_{-18}$\\
$[{\rm Fe/H}]$&Metallicity (dex) 
&$-0.03\pm0.15$&$-0.034\pm0.08$&$0.028^{+0.029}_{-0.028}$&$-0.06^{+0.014}_{-0.013}$&$-0.05\pm0.043$&$0.05\pm0.01$\\
$[{\rm Fe/H}]_{0}$&Initial Metallicity &$0\pm0.13$&$-0.006\pm0.074$&$0.046^{+0.039}_{-0.040}$&$-0.03\pm0.012$&$-0.015\pm0.043$&$0.065\pm0.031$\\
$Age$&Age (Gyr)&$8^{+3.5}_{-3.6}$&$8.2^{+3.3}_{-3.4}$&$7.4^{+3.5}_{-3.1}$&$9.5^{+2.8}_{-3.6}$&$9.3^{+2.8}_{-3.4}$&$7.4^{+3.4}_{-3.0}$\\
 \hline
 \multicolumn{2}{l}{Planet parameters:}\\
 \hline
 $R_{\rm p}$&Planet Radius (R$_{\rm jup}$)&$1.332^{+0.030}_{-0.027}$&$1.331^{+0.026}_{-0.023}$&$1.339^{+0.024}_{-0.023}$&$1.321^{+0.026}_{-0.023}$&$1.323^{+0.023}_{-0.020}$&$1.339^{+0.024}_{-0.022}$\\
 $M_P$&Planet Mass (M$_{\rm jup}$)&$1.185^{+0.055}_{-0.046}$&$1.182^{+0.046}_{-0.038}$&$1.197^{+0.042}_{-0.037}$&$1.164^{+0.048}_{-0.037}$&$1.167^{+0.039}_{-0.031}$&$1.199^{+0.040}_{-0.036}$\\
\hline
\label{tab:globalmodel}
        \end{tabular}
    \end{adjustbox}
\end{table*}

\begin{table}
\centering
\setlength{\tabcolsep}{4pt}
\caption{Transit parameters from the global model.}
        \begin{tabular}{llc}
 \hline \hline
 Symbol & Parameter (Unit) & Value \\
\hline
\multicolumn{2}{l}{Planetary and transit parameters:}\\
\hline
$P_{\rm orb}$&Period (days)&{\scriptsize $1.338230994(84)$}\\
$a$&Semi-major axis (au)&$0.02261^{+0.00036}_{-0.00027}$\\
$i$&Inclination (degrees)&$88.05^{+0.85}_{-0.53}$\\
$T_{\rm eq}$&Equilibrium temperature (K)&$1653\pm12$\\
$R{\rm _P}$/R$_{\star}$&Radius of planet in stellar radii &$0.15205^{+0.00060}_{-0.00062}$\\
$a/R_{\star}$&Semi-major axis in stellar radii &$5.411^{+0.053}_{-0.050}$\\
$\delta$&Fractional transit depth&$0.02312^{+0.00018}_{-0.00019}$\\
$\tau$&Ingress/egress duration (d)&$0.01240^{+0.00029}_{-0.00032}$\\
$T_{14}$&Total transit duration (d)&$0.09027\pm0.00024$\\
$b$&Transit impact parameter &$0.184^{+0.048}_{-0.079}$\\
$u_{1,TESS}$&Linear LD coeff.\ in TESS band &$0.343\pm0.015$\\
$u_{2,TESS}$&Quadratic LD coeff.\ TESS band &$0.253\pm0.022$\\
\hline
\label{tab:transit params}
        \end{tabular}
\end{table}

\subsection{Physical properties without using stellar evolutionary models}
\label{subsec:js_properties}

With the advent of the \textit{Gaia} mission it has become possible to determine the physical properties of the components of transiting planetary systems without needing constraints from stellar evolutionary models. The key additional observable here is the radius of the star, which can be obtained directly from its known distance (via the \textit{Gaia} DR3 parallax) and spectral energy distribution. Once this is available, the properties of the system can be obtained analytically using standard equations from eclipses and celestial mechanics. The process we use is similar to that given in Morrell et al.\ (under review) and can be summarised as follows.

The density of the star can be obtained from only observed quantities and an assumed mass ratio, using the equation
\begin{equation} \label{eq:rhosun}
    \rho_\star = \frac{3\pi}{GP^2}\,\frac{1}{1+q}\,\left(\frac{a}{R_\star}\right)^{3}
\end{equation}
\citep{Roberts99apj,Russell99apj}, where $q = M_{\rm p}/M_\star$ is the mass ratio and $G$ is the gravitational constant. One sets $q=0$ at first, but inserts the known value in subsequent iterations. The stellar mass and surface gravity can then be obtained from its radius and density. The mass of the planet follows from the mass function, and its radius from $R_\star$ and $R_{\rm p}/R_\star$. Its surface gravity, density, and equilibrium temperature can then be determined easily, as can the semimajor axis. Now the masses of both components are known, the $q$ can be inserted in Eq.\,\ref{eq:rhosun} and the calculations repeated. This process takes typically one or two iterations for planetary systems because $q \ll 1$.

For WASP-4 we adopted the radius measurement of $R_\star = 0.912 \pm 0.007$\,R$_\odot$ from \citet{Goswamy++24mn}, who used the infra-red flux method \citep[IRFM;][]{BlackwellShallis77mn} to determine the stellar radii of all of the transiting planetary systems discovered by the SuperWASP survey \citep{Pollacco+06pasp}. We took the $a/R_\star$, $R_{\rm p}/R_\star$ and $i$ from Table~\ref{tab:transit params}, $K_\star = 237.3 \pm 2.2$~m~s$^{-1}$ from \citet{turner2022}, and assumed a circular orbit. The results are given in Table~\ref{tab:js_absdim} and are in good agreement with those in Table~\ref{tab:globalmodel}.

\begin{table} \centering
\caption{\label{tab:js_absdim} Physical properties of the system determined using a direct measurement of the stellar radius from the IRFM.}
\setlength{\tabcolsep}{5pt}
\begin{tabular}{llc} \hline \hline
Symbol & Parameter (Unit) & Value \\
\hline
$M_\star$            & Stellar mass (M$_\odot$)                & $0.899^{+0.033}_{-0.031}$        \\
$R_\star$            & Stellar radius (R$_\odot$)              & $0.912 \pm 0.007$                \\
$\log g_\star$       & Stellar surface gravity (cgs)           & $4.472^{+0.013}_{-0.012}$        \\
$\rho_\star$         & Stellar density ($\rho_\odot$)          & $1.185^{+0.035}_{-0.032}$        \\
$a$                  & Semimajor axis (au)                     & $0.02294^{+0.00028}_{-0.00026}$  \\
$M_{\rm p}$          & Planetary mass (M$_{\rm jup}$)          & $1.200^{+0.032}_{-0.030}$        \\
$R_{\rm p}$          & Planetary radius (R$_{\rm jup}$)        & $1.349^{+0.011}_{-0.012}$        \\
$g_{\rm p}$          & Planetary surface gravity (m\,s$^{-2}$) & $16.33^{+0.38}_{-0.35}$          \\
$\rho_{\rm p}$       & Planetary density ($\rho_{\rm jup}$)    & $0.488^{+0.013}_{-0.011}$        \\
$T_{\rm eq}$         & Equilibrium temperature (K)             & $1672 \pm 28$                    \\
\hline \end{tabular}
\end{table}

\subsection{Transit Timing Analysis}
\label{subsec:ttv_analysis}
\subsubsection{Measurements of the Mid-Transit Timings}
\label{sec:mid_transit_times}
In order to investigate the potential of WASP-4\,b to display significant transit timing variations, we needed to measure all the mid-transit timings of its available light curves to form a homogeneous dataset.  We made use of the web version\footnote{https://exoplanetarchive.ipac.caltech.edu/cgi-bin/ExoFAST/nph-exofast} of the {\sc exofast} modelling suite \citep{eastman2013} to analyze all its selected transit light curves from the Exoplanet Transit Database (ETD)\footnote{http://var2.astro.cz/ETD/}, the relevant literature, and the TESS archive. We specifically used this version ($v1$) of {\sc exofast} because of its easy use and speed with the ameoba fitting algorithm based on the downhill simplex method \citep{neldermead1965}, as well as its advantage in homogeneously selecting the centres and widths of the Gaussian priors for the stellar and planetary fit parameters from previous work selected by the NASA Exoplanet Archive. Parameters for WASP-4\,b were selected from \citet{bouma2019} in our case, which are in very good agreement with our parameters. Uniform priors are used for the linear limb darkening coefficients with the initial values retrieved from the relevant tables in \citet{claret2011} based on stellar atmospheric parameters and the choice of the passband made among the options provided by a drop-down menu. We selected the I-band for the TESS passband because their transmission functions are similar. Orbital periods are fixed to the literature value because a single light curve normalized to maximum light is modelled in each of the runs. The mid-transit time, set to be free at the beginning, is determined from the contact points of the light curve model and provided by the program together with its uncertainty and other model parameters, such as the full and total transit durations and the transit depth, which we used to double-check the consistency of the results and their agreement with system parameters. 

\begin{figure}
    \centering
    \includegraphics[width=\columnwidth]{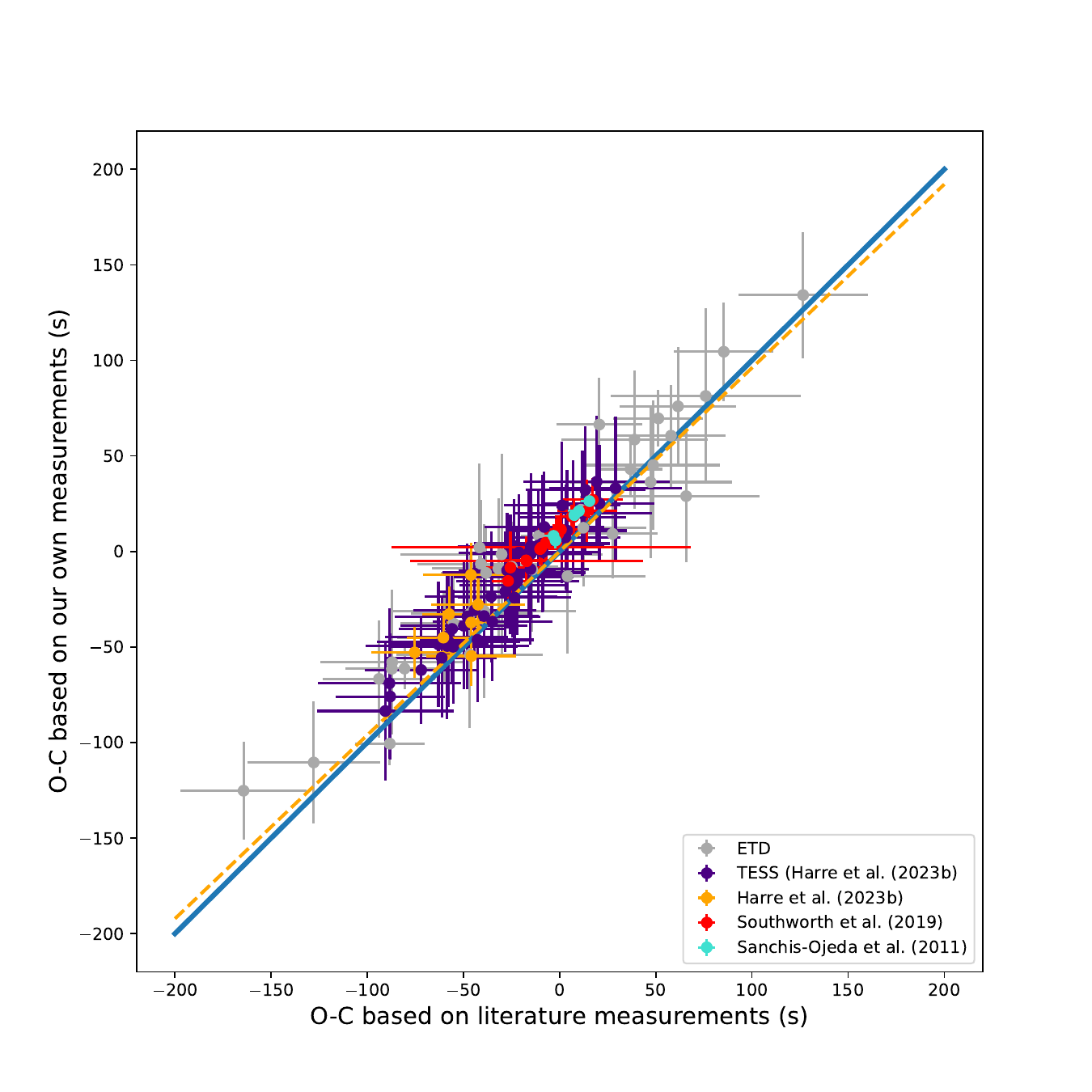}
    \caption{Comparison of our own measurements of mid-transit times and those from the literature based on the O-C values computed by employing linear ephemerides provided by \citet{southworth2019}, mid-transit times from the literature ($x$-axis), and those from our own measurements with the {\sc exofast} code ($y$-axis) in seconds. The blue line is for equal O-C's, while the dashed orange line is the best linear fit.}
    \label{fig:measurement_comparison_wasp4}
\end{figure}

We were unable to obtain the light curves of all transit observations in the literature from which mid-transit measurements were made and published. Therefore, we needed to check if our measurements of mid-transit timings agree with those in the literature in general. This data set includes 36 ETD, 46 TESS light curves, the mid-transit times of which were published by \citet{turner2022}, 7 light curves from \citet{sojeda2011}, 15 from \citet{southworth2019}, and 5 from \citet{harre2023b}, adding up to 109 light curves in total. In order to compare two sets of measurements, we first checked if there are measurements that do not agree with each other to within the 1$\sigma$ level, and we found no such measurements. While the average of the measurement error published by other authors is 27.48 seconds, and that in our measurements from the same light curves is 25.75 seconds, the average difference between the measurements from the same light curves is only 11.83 seconds. However, we measured a larger mid-transit time value in 98 out of these 109 light curves, which indicates a positive shift in our measurements with respect to that appeared in the literature and the ETD. When we repeat this experiment only with TESS data, which are both in BJD$_{\rm TDB}$ timing frame, 43 of 46 measurements turned out to indicate a later mid-transit time in our measurements compared to that of \citet{turner2022}. Although we were unable to precisely identify the root cause of this offset; we suspect it arises from differences in the  modelling and mid-transit time measurement approaches employed by these studies and those in use by {\sc exofast} because \citet{turner2022} also analyzed the DVT light curve product and the average difference between the measurements from the same light curves is 10.87 seconds. We then compared the differences between the observed (O) and calculated (C) mid-transit times, based on the linear ephemerides provided by \citet{southworth2019}, by plotting both sets of measurements against each other (Fig. \ref{fig:measurement_comparison_wasp4}). For clarity, we included error bars for both measurements in seconds. The two sets of measurements show good agreement, as indicated by the slope of the best-fit linear relation (dashed orange line in Fig. \ref{fig:measurement_comparison_wasp4}), $m = 0.981 \pm 0.020$, which is consistent with unity (i.e. equal O–C values, shown by the continuous blue line) within 1$\sigma$. However, the systematic offset in our mid-transit times is reflected in the $y$-intercept of the fit ($11.13$ s), which is nearly equal to the average difference between the two data sets ($11.83$ s). Consequently, we proceeded with our TTV analysis using both sets of mid-transit time measurements separately.

\subsubsection{Construction and Analysis of the TTV Diagram}
\label{sec:ttv_diagram}
We used the reference linear ephemeris elements provided by \citet{southworth2019} to construct the TTV diagram, which we then fit with a linear model, a quadratic model, and an apsidal motion model. In each of the cases, we have sampled from the Gaussian priors for model parameters centred at values determined by a non-linear least squares fit by using its uncertainties as widths. We sampled from the prior distributions making use of 16 Markov chains. We then calculated the log-likelihood functions of each draw and formed the posterior distributions of the fit parameters, which we sampled from at the end, throwing away the first chunks as the burn-in period. We used the current ($5^{\rm th}$) version of {\sc PyMC}\footnote{https://www.pymc.io/welcome.html} \citep{pymc2023} for the probabilistic fitting procedure and {\sc lmfit} for the preliminary non-linear least squares fitting \citep{lmfit2016}. We provide the parameters of the probabilistic fitting procedure and the results we obtained for each of the cases using both data sets in Table~\ref{tab:ttv}.

\begin{table}
\centering
\caption{Results of the TTV Analysis}
\label{tab:ttv}
\begin{tabular}{lcc}  
\hline
\hline
Parameter & All data & Our Measurements \\
\hline
\multicolumn{3}{l}{Linear Model ($m \times E + n$)}\\
\hline
Number of Chains & 16 & 16\\
Number of Draws & 50000 & 50000\\
Burn-in Period & 5000 & 5000 \\
Slope (m $\times 10^{-7}$) & $-1.67^{+0.14}_{-0.14}$ & $-1.86 ^{+0.16}_{-0.16}$ \\
Intercept (n $\times 10^{-5}$) & $0.30^{+2.03}_{-2.04}$ & $-1.39 ^{+2.58}_{-2.58}$ \\
$\chi^2_{\nu}$ & 4.69 & 5.00 \\
AIC & 335.62 & 256.16 \\
BIC & 342.38 & 262.29 \\
\hline
\multicolumn{3}{l}{Quadratic Model ($a \times E^2 + b \times E + c$)}\\
\hline
Number of Chains & 16 & 16\\
Number of Draws & 50000 & 50000\\
Burn-in Period & 5000 & 5000 \\
Quad.Coeff. (a $\times 10^{-11}$) & $-9.81^{+1.21}_{-1.21}$ & $-8.52 ^{+1.47}_{-1.47}$ \\
Lin.Coeff (b $\times 10^{-8}$) & $-4.86^{+1.92}_{-1.92}$ & $-8.37 ^{+2.29}_{-2.28}$ \\
Intercept (c $\times 10^{-4}$) & $1.84 ^{+0.29}_{-0.29}$ & $1.91 ^{+0.39}_{-0.39}$ \\
$\chi^2_{\nu}$ & 3.61 & 4.14 \\
AIC & 280.29 & 227.55 \\
BIC & 290.42 & 236.74 \\
\hline
\multicolumn{3}{l}{Apsidal Motion Model}\\
\hline
Number of Chains & 16 & 16\\
Number of Draws & 100000 & 100000\\
Burn-in Period & 25000 & 25000 \\
Eccentricity (e) & $0.0020^{+0.0007}_{-0.0009}$  & $0.0013^{+0.0005}_{-0.0009}$ \\
Arg. of Periastron ($\omega ~[^{\circ}$]) & $2.13^{+0.17}_{-0.23}$ & $1.53^{+0.15}_{-0.17}$ \\
$\omega dE$ ($\times 10^{-4}~^{\circ}$/cycle) & $5.67^{+0.98}_{-1.50}$ & $7.93^{+1.75}_{-2.26}$ \\
Slope (m$_0$ $ \times 10^{-7}$) & $-3.93 ^{+1.20}_{-0.96}$ & $-4.55 ^{+1.59}_{-1.09}$ \\
Intercept (n$_0$ $ \times 10^{-4}$) & $-2.69 ^{+2.62}_{-1.75}$ & $9.42 ^{+7.35}_{-3.91}$ \\
$\chi^2_{\nu}$ & 11.46 & 8.75 \\
AIC & 2428.66 & 1349.32 \\
BIC & 2445.54 & 1364.63 \\
\hline
\hline
\end{tabular}
\end{table}

For both data sets, which are based on all data (column 1 in Table-\ref{tab:ttv}) and only the light curves that we modelled and measured the mid-transit times from (column 2 of the same table), the quadratic model turned out to be superior to both the linear and apsidal motion models in all metrics of goodness of fit. With reduced-$\chi^2$ values of 3.61 \& 4.14, the quadratic model with a negative quadratic coefficient is the best model for both data sets. In addition, the quadratic model performed significantly better than the linear model in the representation of the TTV data in terms of Akaike (AIC) \citep{akaike1974} and Bayesian Information Criteria (BIC) \citep{schwarz1978,liddle2007}, as given below for all data including the literature minima and the mid-transit times we measured ourselves, respectively.\\ 

$\Delta \mbox{AIC} = \mbox{AIC}_{\rm quad} - \mbox{AIC}_{\rm lin} = -55.33 \mbox{~~~and~~~} -28.61$ \\ 

$\Delta \mbox{BIC} = \mbox{BIC}_{\rm quad} - \mbox{BIC}_{\rm lin} = -51.96 \mbox{~~~and~~~} -25.55$ \\

Since both data sets ended up with similar fit statistics and there is agreement between them as shown in Fig. \ref{fig:measurement_comparison_wasp4}, we decided to continue with the results based on the entire data set so as not to discard observations that are not available to us. We analyzed ETD data separately due to their large scatter incompatible with their uncertainties. Both models represent the data with similar fit statistics. The large $\chi^2_{\nu} \sim~8.00$ show that the error bars of these measurements are underestimated. When these data are removed from the full data set, and the TTV-fits are updated, $\chi^2_{\nu}$ of the linear and quadratic fits are found to improve to the values of 4.01 \& 2.53, and $\Delta \mbox{AIC} = 170.14 - 252.58 = -82.44$ \&  $\Delta \mbox{BIC} = 179.71 - 258.97 = -79.26$. Nonetheless, we continue with the full data set for completeness. Since we do not have a-priori information on the source of the large scatter in the mid-transit timings that lead to rather high $\chi^2_{\nu}$ values in the fits, we did not remove any data points based on a tighter constraint on the sigma-outliers than the current value of $3\sigma$. However, we suspect that both the underestimation of the error bars, most probably in photometry propagated to the timing measurements, and the spot-induced asymmetries play a role in the scatter of the timing measurements on the TTV diagram being larger than the error bars suggest.

As a result, our linear model ended up with the following linear ephemeris elements (Eq.\,\ref{eq:lin_ephemeris}) and a $\chi^2_{\nu}$ value of 4.69.

\begin{equation}
  T = 2456505.748956(23) + 1.338231262(14) \times E
  \label{eq:lin_ephemeris}
\end{equation}

while the quadratic ephemerides with a $\chi^2_{\nu}$ value of 3.61 are given with Eq-\ref{eq:quad_ephemeris}.

\begin{equation}
\begin{split}
  T = 2456505.749138(29) + 1.338231380(19) \times E \\
  - 0.98(12) \times 10^{-10} \times E^2
\end{split}
  \label{eq:quad_ephemeris}
\end{equation}

\begin{figure*}
\includegraphics[width=0.9\paperwidth]{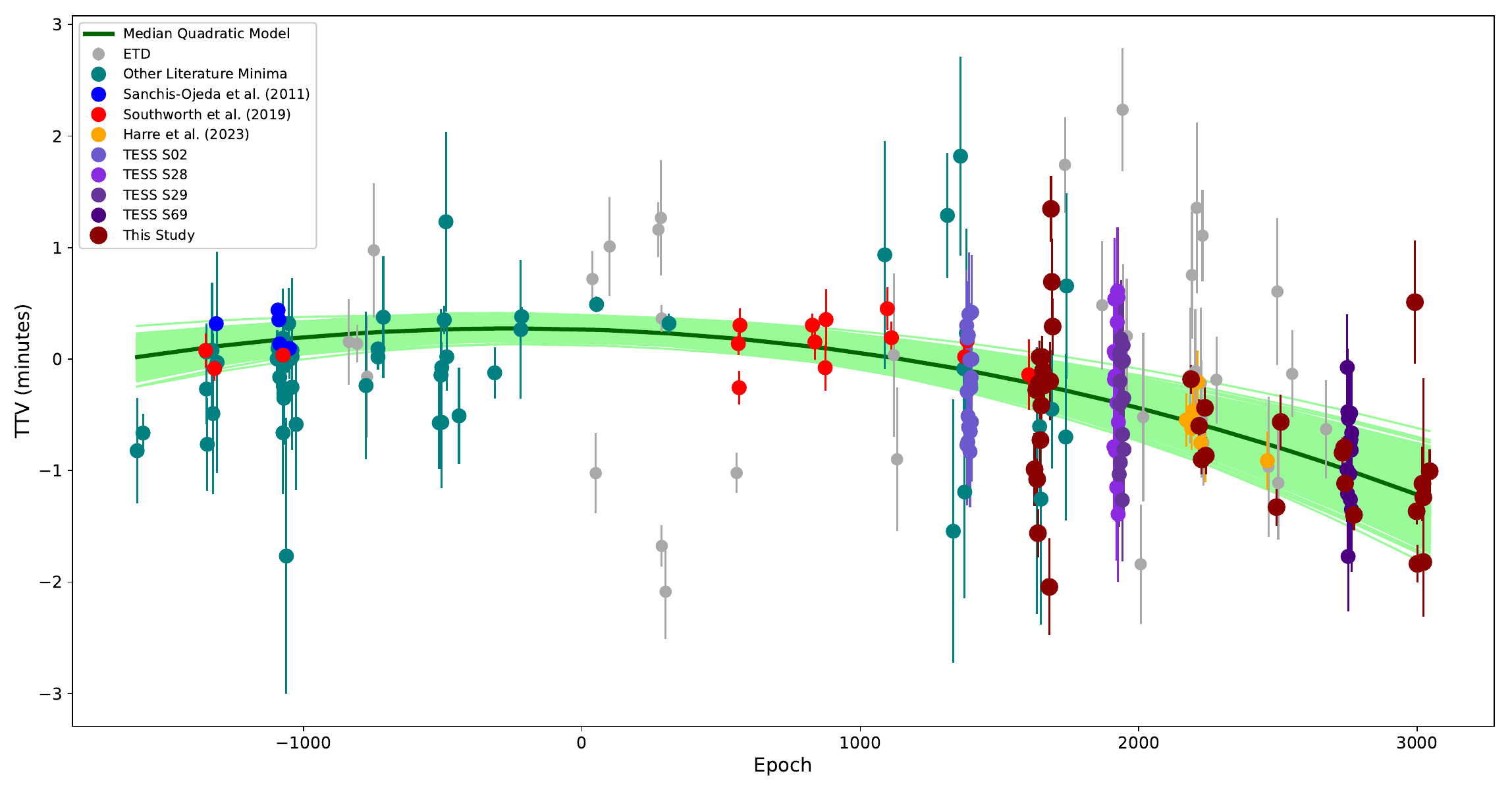}
\caption{TTV diagram of WASP-4\,b with the strongly favoured quadratic model superimposed on the timing data. The last 5000 models have been plotted with the lighter shade of green, while the model based on the median values of model parameters from their posterior distributions is provided in dark green.}
\label{fig:wasp4_ttv_diagram}
\end{figure*}

These results strongly favour the quadratic model with a Bayes factor of B$_{ql} = e^{-\Delta \mbox{BIC} / 2} = 1.92 \times 10^{11}$ \citep{masson2011}. Therefore, we provide only the quadratic model and plotted on the entire dataset, including mid-transit times from the literature, in Fig. \ref{fig:wasp4_ttv_diagram}.\\

The tidal quality factor based on the median value of the quadratic coefficient of our quadratic model as given by \citet{goldreich1966, patra2017} is
\begin{equation}
    Q^{\prime}_{\star}= -\frac{27\pi}{2} \left(\frac{M_{\rm p}}{M_{\star}}\right) \left(\frac{R_{\star}}{a}\right)^{5}\frac{1}{\Dot{P}} = (8.08 \pm 1.17) \times 10^{4}\\
\end{equation}

for the stellar parameters derived from our Model-4. Since the primary factor that affects $Q^{\prime}_{\star}$ is the quadratic coefficient of the TTV model, different values for the masses of the planet and its host star only change this value between 79405 and 80871 for the different models in Table \ref{tab:globalmodel}. These agree with each other to within 1$\sigma$ of their respective uncertainties.

We continued our analysis with the residuals of the O-C values from the linear model to search for potential periodic signals in the data, which could hint at the existence of gravitationally bound perturbers. We found a statistically significant periodicity at $387.96$ days with a False Alarm Probability value of $2.48 \times 10^{-8}$. However, when we phase-fold the TTV diagram, we did not observe any variation in the data. Fitting a Keplerian with a fixed periodicity to this value ends up within much poorer fit statistics compared to linear and quadratic models. If this signal was caused by a perturber, it should have been detected in the RV signal too. The 7001.0-day periodicity found in the RV data by \citet{turner2022}, on the other hand, does not show up in our frequency search. When we fix the period to this value and obtain the best-fitting Keplerian, the amplitude turns out to be 19.64 seconds, which requires a very massive object to orbit the system in such a long-period orbit. The planetary-mass body suggested by \citet{turner2022} would lead to a much smaller TTV-amplitude, which cannot be detected with the precision of the mid-transit times.

Since the stellar rotation was found to be $P_\mathrm{rot}= 22.2 \pm 3.3$ days by \citet{2017A&A...602A.107B}, we attempted to restrict the frequency interval and update our frequency search due to the clear asymmetries on some of the transit profiles to investigate if other peaks showing up in the Lomb-Scargle periodogram are caused by the magnetic activity. These asymmetries, on the other hand, did not allow us to measure a rotation rate for the host star because the spots causing them are not persistent on the transit chord in the timescales covered by neither precise ground-based light curves of ours and others, nor TESS light curves. When we restrict the frequency successively to 0.01 and 0.05 cycles per day, we obtain periodicities at 51.03 and 15.74 days, with FAP values of $6.62 \times 10^{-4}$ and $1.79 \times 10^{-3}$, respectively. When we perform a frequency analysis with a dataset based only on our observations, the 51-day periodicity disappears from the periodogram, while the $\sim15$-day and $\sim386$-day peaks shift slightly to longer periods with higher FAP values. Furthermore, we repeated the analysis with a dataset excluding the ETD-transits and we observed peaks at different frequencies. Phase-folding TTV diagrams based on any of these frequencies does not provide evidence for a periodic behaviour in the TTV diagram. Therefore, we conclude that we do not have evidence for neither a perturber nor a cyclic magnetic activity in our frequency analyses.

We also modelled the TTV data based on an apsidal motion model in the same manner, although we have fixed the eccentricity to zero in our global models. Since we haven't fit the RV data, we made use of the values of eccentricity, argument of periastron ($\omega$), and its time derivative ($d\omega / dE$) found by \citet{bouma2019} as Gaussian priors for these parameters. Although the median values of the posterior distributions are in very good agreement with the findings of \citet{bouma2019}, the AIC and BIC values do not favor this model, compared to the linear or quadratic models. Since the tidal circularization timescale is expected to be very short for this planet's orbit and we do not have any new RV data or occultation observations to suggest a non-zero eccentricity, we do not argue in favour of this apsidal motion model no matter how we restrict our data sets to cover literature and / or ETD mid-transit times, although we cannot rule it out either.

The acceleration of the system with respect to Earth at a rate of 0.0422 m s$^{-1}$ day$^{-1}$ has been suggested to explain the observed quadratic change in the TTV data by \citet{bouma2020} based on their RV model. However, this hypothesis has been conclusively ruled out later by both \citet{baluev2020} and \citet{turner2022}, who modelled new RV observations of the target together with previous data. This is why we haven't included an acceleration term in our fits.

\section{Discussion}
\label{sec:discussion}

To explore whether the inferred $Q'_\star$ is compatible with tidal theory \citep[e.g.][]{B2020,2024MNRAS.527.5131B}, we have constructed stellar models of WASP-4 consistent with the parameters in Tables~\ref{tab:globalmodel} and \ref{tab:transit params} using {\sc MESA} version r24.08.1 \citep{Paxton2011,Paxton2013,Paxton2015,Paxton2018,Paxton2019,Jermyn2023}, with stellar parameters from MIST \citep{MIST02016,choi2016}. This provides us with radial profiles ($r$ is the spherical radius from the stellar centre) for the stellar density $\rho(r)$, pressure $p(r)$, gravitational acceleration $g(r)$, and other variables, as a function of stellar age. Using these profiles, we can calculate the tidal response.

\begin{figure*}
\centering
 \begin{subfigure}{0.33\textwidth}
 \centering
 \includegraphics[trim=7cm 0cm 8cm 0cm,clip=true,width=0.8\textwidth]{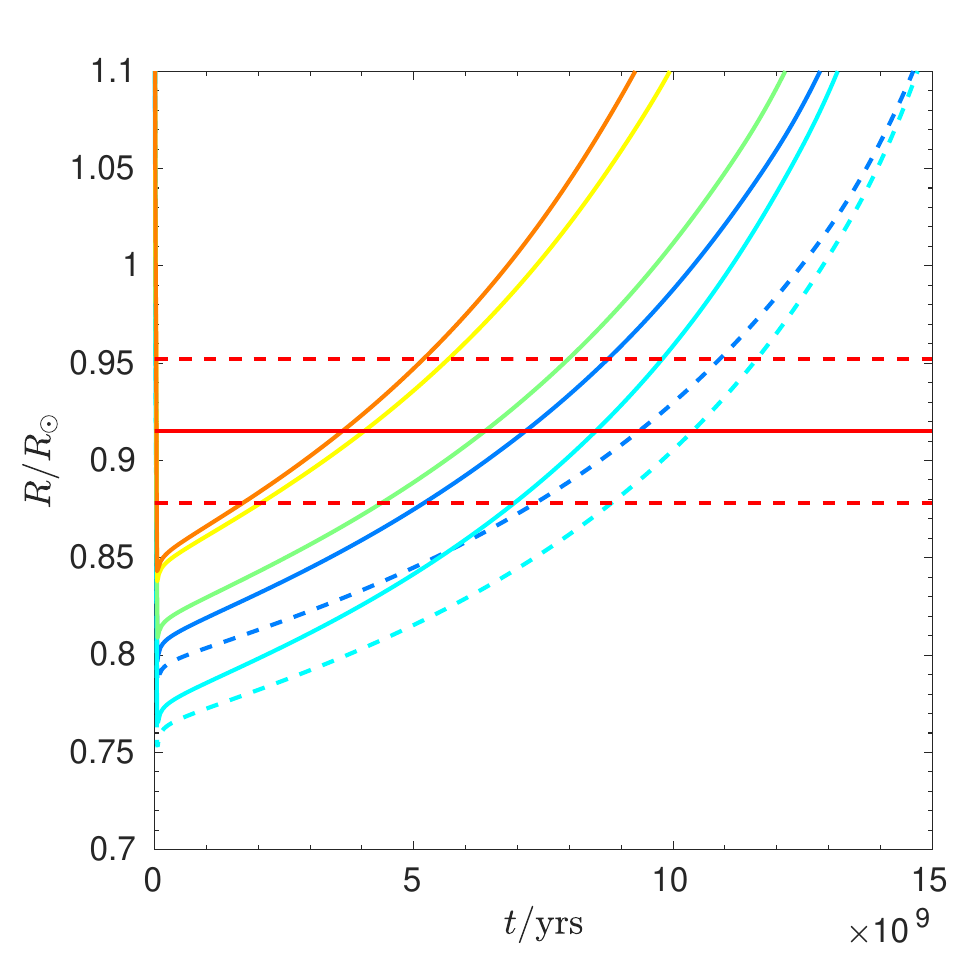}
 \end{subfigure}
 \begin{subfigure}{0.33\textwidth}
 \centering
  \includegraphics[trim=7cm 0cm 8cm 0cm,clip=true,width=0.8\textwidth]{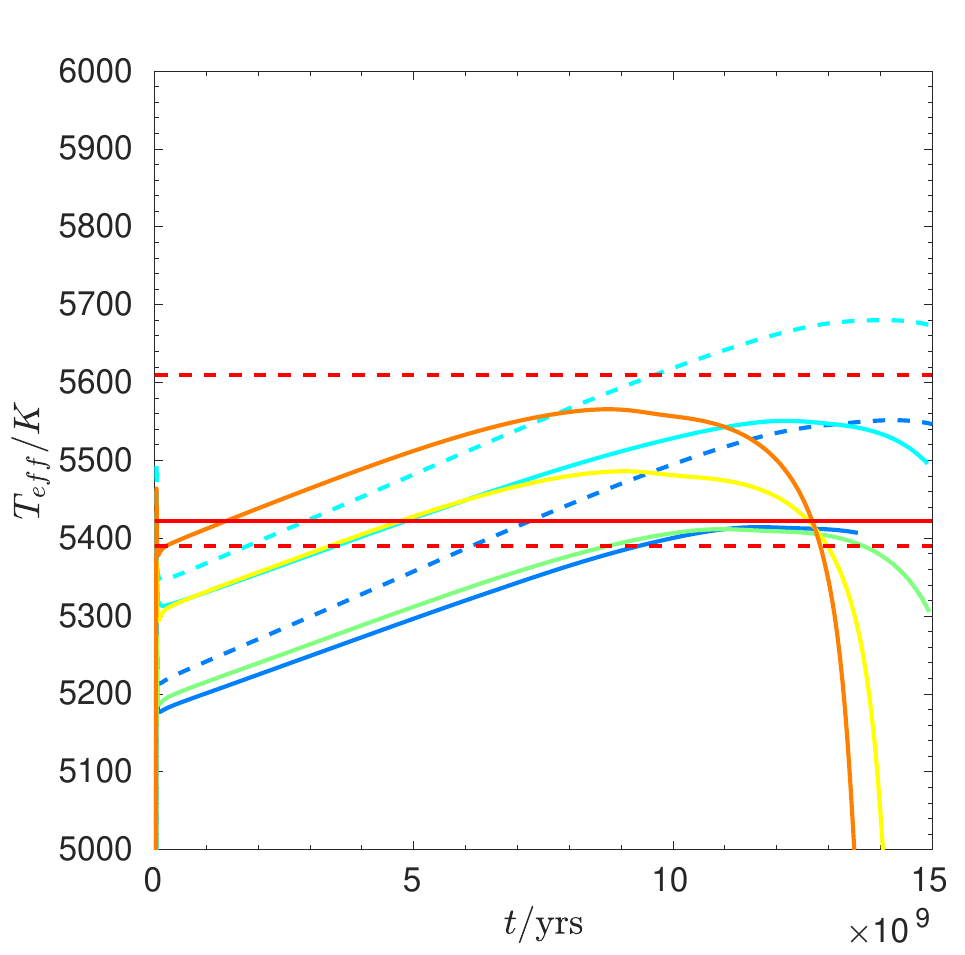}
  \end{subfigure}
  \begin{subfigure}{0.33\textwidth}
  \hspace{0.5cm}
  \includegraphics[trim=17.5cm 8cm 3.5cm 1cm,clip=true,width=0.8\textwidth]{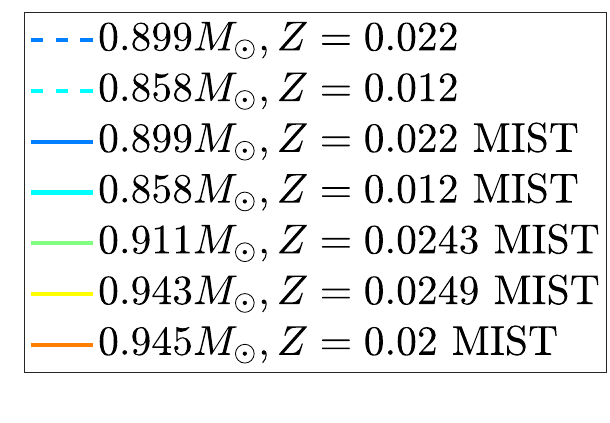}
  \end{subfigure}\\
  \begin{subfigure}{0.33\textwidth}
 \centering
   \includegraphics[trim=7cm 0cm 8cm 0cm,clip=true,width=0.8\textwidth]{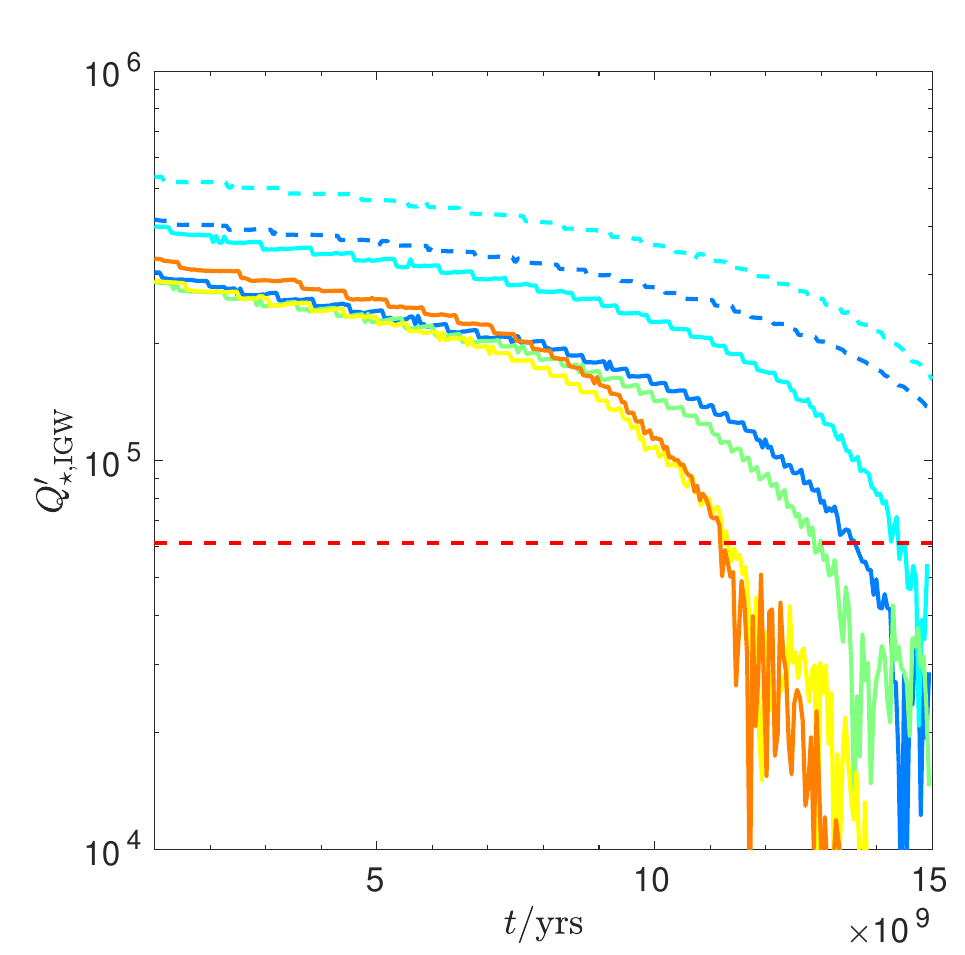}
   \end{subfigure}
   \begin{subfigure}{0.33\textwidth}
 \centering
   \includegraphics[trim=7cm 0cm 8cm 0cm,clip=true,clip=true,width=0.8\textwidth]{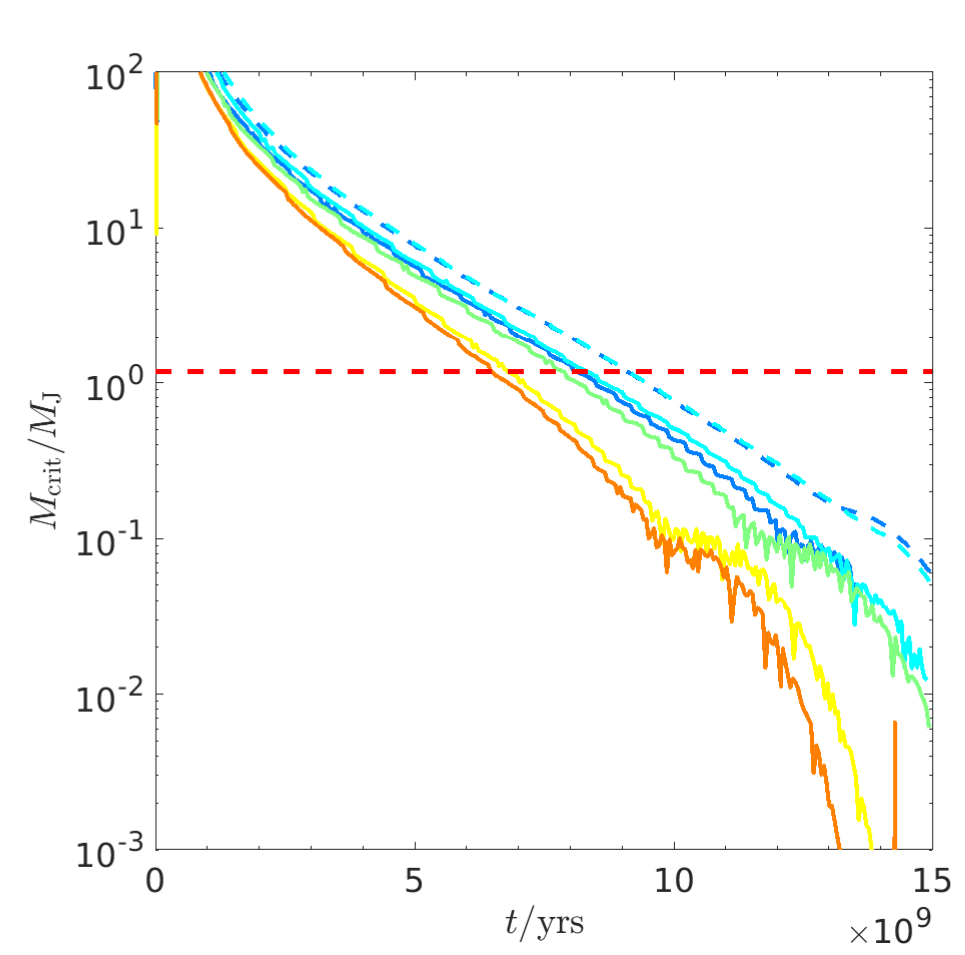}
    \end{subfigure}
     \begin{subfigure}{0.33\textwidth}
 \centering
 \includegraphics[trim=7cm 0cm 8cm 0cm,clip=true,clip=true,width=0.8\textwidth]{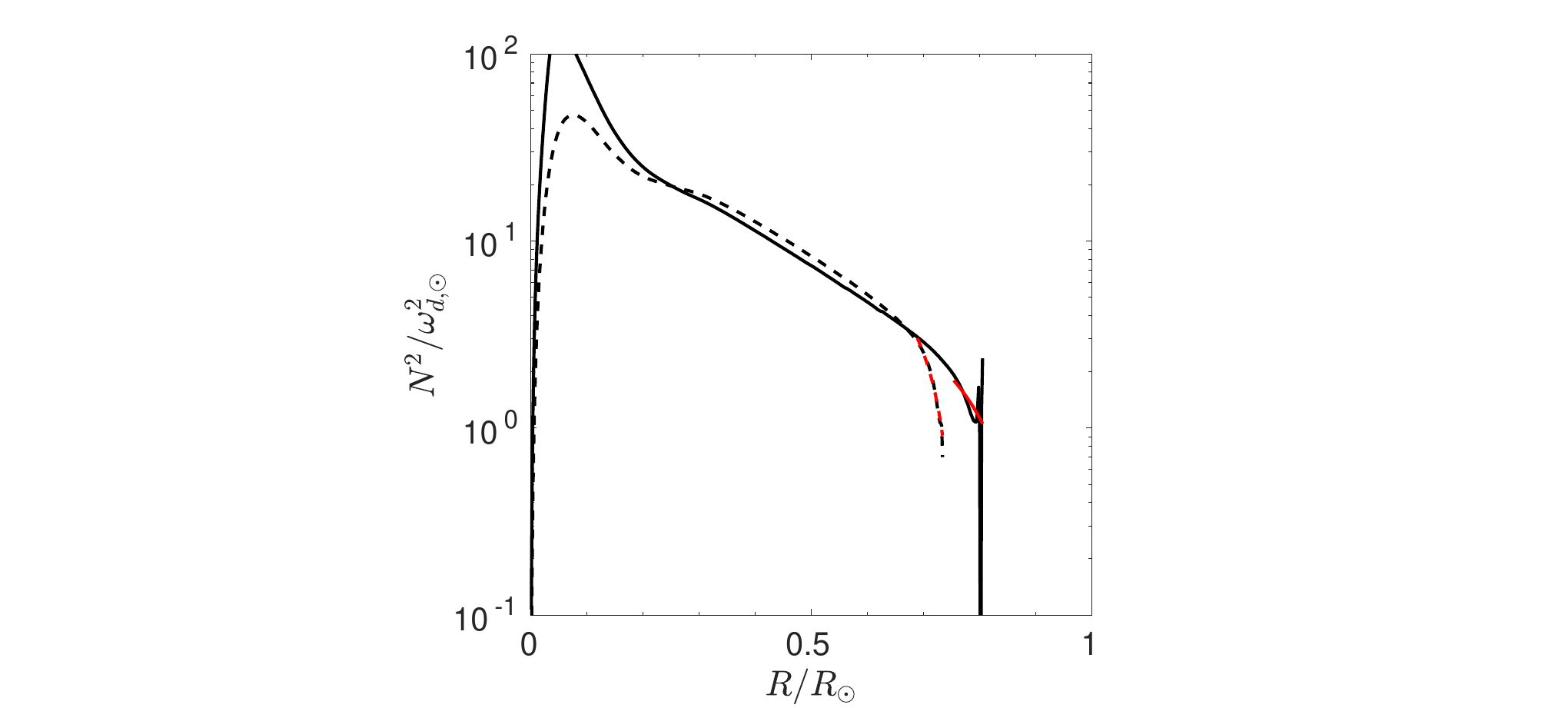}
     \end{subfigure}
  \caption{Stellar properties and tidal quality factor computed using either {\sc MESA} default parameters or MIST parameters for the initial masses and metallicities specified in the legend. Panels (a) and (b) show the stellar normalised radius and effective temperature, over-plotting the observational constraints as horizontal red dashed lines based on the maximum reported deviations in Table 2 (the solid red line is the mean value of model 4). Panel (c) shows the modified tidal quality factor $Q^{\,\prime}$ for internal gravity waves (with the observational constraint as the red dashed line) and panel (d) indicates the critical planetary mass required for wave breaking to be predicted in the stellar core (with the planetary mass as the red dashed line). Panel (f) shows the radial profile of the Brunt-V\"ais\"a ll\" a frequency (normalised by the solar squared dynamical frequency $\omega_{d,\odot}^2=GM_\odot/R_\odot^3$) in the model with $M=0.945M_\odot$ and $Z=0.02$ at two times ($t=8.22$ Gyr in dashed lines and $t=10.7$ Gyr in solid ones, showing a local linear fit employed in calculating $Q'_\mathrm{\star,IGW}$).}
  \label{AB1}
\end{figure*}

Since the star is slowly rotating, inertial waves cannot be excited inside the star by a planet on an aligned orbit (for which a stellar rotation period shorter than 2.67 d would be required); we have verified that equilibrium tides are also ineffective, according to our current understanding of their dissipation \citep[e.g.][]{DBJ2020,B2020}. Theoretical models predict the excitation and dissipation of internal gravity waves in the radiative core to be the dominant tidal mechanism for any reasonable stellar model. These waves are launched inward from the radiative/convective interface, and if they are fully damped \citep[either by wave breaking, critical layer absorption or magnetic wave conversion e.g.~][]{BO2010,Barker2011,Guo2023,Weinberg2024,Duguid2024}, the resulting $Q'_\star$ can be straightforwardly computed using the stellar interior profiles. To do so, we calculate the dissipative tidal response, including $Q'_\star$, by evaluating Eq.~41 of \citet{B2020} \citep[and the surrounding formalism; see also][]{GD1998,Chernov2017,Ahuir2021,MaFuller2021}. We show our results in Fig.~\ref{AB1}, with the predicted values of $Q'_\star$ due to gravity waves (with a subscript IGW) in various stellar models plotted in the bottom left panel.

We find $Q'_\star\approx 2-5\times 10^5 \left(P_\mathrm{tide}/0.67\,\mathrm{d}\right)^{8/3}$ (evaluated using the tidal period $P_\mathrm{tide}=(P_\mathrm{orb}^{-1}-P_\mathrm{rot}^{-1})^{-1}/2\approx P_\mathrm{orb}/2$, which takes the value $0.67$ d here) 
in our main sequence models for $t\leq 8$ Gyr \citep[in agreement with the values reported in][]{B2020}. On the other hand, towards the end of the main sequence ($t\gtrsim 10$ Gyr), smaller values compatible with the observed one ($Q'_\star\approx 6\times 10^4$) can be obtained, though only for a brief period in the evolution, depending on stellar mass and metallicity. We have explored models with a variety of stellar masses and initial metallicities within the observational 1$\sigma$ error bars (in Table~\ref{tab:globalmodel}), and models including either default {\sc MESA} parameters or MIST ones, to determine the stellar parameters for which $Q'_\star\approx 6\times 10^4$ can be achieved within 13 Gyr. We find that this is possible but only in our models at the more massive end in Table~\ref{tab:globalmodel}, with $M_\star\gtrsim 0.911$ M$_\odot$ (and with initial metallicity $Z\approx 0.02$ or larger here, but this appears to be less crucial). However, the stellar radius is larger than the observational constraints in these models when $Q'_\star\approx 8\times 10^4$ (top left panel of Fig.~\ref{AB1}; the effective temperature is plotted in the middle top panel), so these models may not be representative of WASP-4.

In the bottom right panel, we show illustrative radial profiles of the Brunt-V\"ais\"all\" a frequency $N^2$ (normalised by the square of the solar dynamical frequency $\omega_{d,\odot}^2=GM_\odot/R_\odot^3$) in the model with $M=0.945M_\odot$ and $Z=0.02$ at two different times ($t=8.22$ Gyr in dashed lines and $t=10.7$ Gyr in solid ones). Near the interface, a bump in $N^2$ appears for later ages, which is sensitive to the stellar mass, metallicity, and the choice of semiconvective mixing and elemental diffusion parameterisations. Since the latter are particularly uncertain, the appropriate value for the radial gradient of $N^2$ near the interface (the mean value over the first gravity wavelength is necessary to compute $Q'_\mathrm{\star,IGW}$, for which we plot the local linear fits we employed in red lines) leads to some uncertainty in predictions for $Q'_\star$ for a given stellar mass \citep[see also][]{Barker2011}. However, the shallower slope and larger radius of the radiative/convective interface in the older model here (partly compensated by the density varying there by 30\%) explain why $Q'_\mathrm{\star,IGW}$ changes from $1.83\times 10^5$ to $8.32\times 10^4$ between these two ages \citep[see Eq.~41 of][]{B2020}.

In the bottom middle panel of Fig.~\ref{AB1}, we show the critical planetary mass required for wave breaking in the stellar core. WASP-4b's mass exceeds this critical value in all models before 10 Gyr. Hence, it is reasonable to expect the gravity waves to be in the fully damped regime \citep{BO2010,Barker2011,Guo2023}, thereby justifying our approach to calculate $Q'_\star$.

Another possibility we have considered, if the star is younger, could be for the tide to resonantly excite a g-mode, leading to enhanced dissipation compared with the fully damped regime as long as the waves remain in the linear regime. Since the planet is very likely to cause wave breaking in the stellar core (hence justifying the fully damped regime) in all modes with ages older than $7-10$ Gyr, the only way the tide can remain in the linear regime when a resonance is passed through by the system is if the star is younger (when $M_\mathrm{crit}$ is larger). This is because passing through a resonance would enhance the wave amplitude and therefore require a lower planetary mass for wave breaking compared with that predicted by the solid lines in the $M_\mathrm{crit}$ figure. We have briefly explored this possibility using {\sc GYRE} \citep[e.g.][]{GYRE2013,GYRE2023}. We find that enhanced dissipation is only possible extremely close to resonances with g-modes, and these have very narrow widths in frequency (because the g-modes are not particularly high radial order ones, so radiative damping is weak). Dissipation between resonances is much weaker, leading to $Q'_\star\gtrsim 10^9$ for most frequencies; thus we do not view this possibility as a particularly likely one. Finally, we have explored whether ``resonance locking" could in principle work, where stellar evolution could maintain the system in a g-mode resonance with enhanced dissipation. Using Eq. 14 of \citet{MaFuller2021} gives $Q'_\star\gtrsim 10^7$ for this mechanism, so it is unlikely to be relevant here \citep[see also some theoretical objections in][for solar-type stars]{Guo2023}.

In summary, we find that it is possible to theoretically match the observed $Q'_\star\approx 8\times 10^4$ by gravity wave damping in the stellar core, but only in older stars with ages older than approximately 11 Gyr, or in models that do not quite match the inferred radius of WASP-4. Otherwise, values of $Q'_\star$ that are approximately 2-4 times larger are predicted in most models. It is likely that further tweaks to the stellar model parameters or input physics could provide a better match with the observed value while also satisfying the observational constraints on radius and effective temperature. So we conclude that gravity wave damping in the stellar core is a possible way to explain WASP-4\,b's orbital decay, with the above caveats.
 
\section{Conclusion}
\label{sec:conclusion}
WASP-4\,b is an interesting hot-Jupiter whose orbital period has been decreasing. The deviation from the linear ephemeris has been attributed to various factors, including the system's line-of-sight acceleration \citep{bouma2020}, the influence of an external perturber \citep{turner2022}, at least contributing to the orbital period variation via light-time effect \citep{harre2023b}, and a potential apsidal motion \citep{harre2023a}. 

In this study, we compiled and analyzed 37 complete transit light curves of WASP-4\,b from the literature, excluding seven from our transit timing variation (TTV) analysis for various reasons. Additionally, we examined 41 light curves from the Exoplanet Transit Database (ETD), with five eliminated from further analysis. We also processed a total of 61 light curves acquired by TESS in four sectors at a two-minute cadence, and incorporated 37 newly obtained observations with three different telescopes, of which three were excluded (Appendix \ref{app:lightcurves}). In total, we analyzed 158 light curves and homogeneously determined mid-transit timings using {\sc exofast} models, forming the most extensive TTV dataset for WASP-4b to date. Furthermore, we included 58 mid-transit timings from the literature derived from light curves that were not directly accessible. As a result, our final TTV diagram consists of 216 data points.

We analyzed the TTV diagram using linear, quadratic, and apsidal motion models. The quadratic model proved to be statistically superior, even when excluding mid-transit timings from the literature that were not derived from our own light curve models. Additionally, we did not detect any statistically significant periodic signals in the data. Consequently, we adopted the quadratic model and used the inferred rate of period change to estimate the reduced tidal quality factor (Q$^{\prime}_{\star}$), assuming that this change is caused by tides in the star.

To assess the significance of our derived value of Q$^{\prime}_{\star} \sim 80000$, we required a global model from which we extracted the system parameters. We then used these parameters to construct theoretical models of WASP-4 and its dissipative tidal response. A main-sequence star with the fundamental properties of WASP-4 is found to be insufficiently dissipative to explain the observed value; instead, we predict Q$^{\prime}_{\star}\sim 2 - 5 \times 10^5$ due to the dissipation of internal gravity waves in its radiative core (the most effective tidal mechanism we studied) assuming that these waves are fully damped. Given the system's estimated age ($< 10$ Gyrs) we derived from isochrones, we expect gravity waves to be in the fully damped regime due to wave breaking in the stellar core. Some of our models are able to explain the relatively stronger dissipation observed, but they require a more massive star ($M_{\star} > 0.911$ $M_{\odot}$), with slightly higher metallicity, at the end of its main-sequence life. However, the stellar radius in these models is larger than that observed for WASP-4. Further tweaks to the stellar model parameters and input physics may be able to produce models that provide a better match to observations. We also argue that resonance locking, or the system being in close proximity to resonance with a g-mode oscillation, are unlikely to be able to enhance the dissipation to the observed level (though the latter is not definitively ruled out).

At present, orbital decay remains the only viable explanation for the observed period change. However, its inferred rate suggests that a more precise understanding of the host star parameters is necessary to reconcile it with current tidal theory. Only the most massive star models compatible with our constraints, in the later stages of its evolution, appear to be able to dissipate tidal energy efficiently enough to explain the observed Q$^{\prime}_{\star}$. In this respect, WASP-4\,b presents a similar case to WASP-12\,b \citep{ack2023} for which the orbital decay scenario is the only leading explanation for its TTVs, and its host star parameters and evolutionary stage are under debate \citep[e.g.][]{yee2020}. Both targets deserve more investigation in terms of stellar modeling to be able to explain the observed orbital decay within the current framework of  tidal interaction theory.

\section*{Acknowledgments}

We gratefully acknowledge the support by The Scientific and Technological Research Council of Turkey (T\"UB\.{I}TAK) with the project 123F293. AJB was partially funded by STFC grant ST/W000873/1. JS was partially funded by STFC grant ST/Y002563/1. Based on data collected by MiNDSTEp with the Danish 1.54\,m telescope at the ESO La Silla Observatory. Some of the data presented in this paper were obtained from the Multimission Archive at the Space Telescope Science Institute (MAST). STScI is operated by the Association of Universities for Research in Astronomy, Inc., under NASA contract NAS5-26555. Support for MAST for non-HST data is provided by the NASA Office of Space Science via grant NAG5-7584 and by other grants and contracts. This research has used the NASA Exoplanet Archive, which is operated by the California Institute of Technology, under contract with the National Aeronautics and Space Administration under the Exoplanet Exploration Program. This work presents results from the European Space Agency (ESA) space mission Gaia. Gaia data are being processed by the Gaia Data Processing and Analysis Consortium (DPAC). Funding for the DPAC is provided by national institutions, in particular the institutions participating in the Gaia MultiLateral Agreement (MLA). We thank all the observers who report their observations to Exoplanet Transit Database (ETD).
 TRAPPIST-South is funded by the Belgian Fund for Scientific Research (Fond National de la Recherche Scientifique, FNRS) under the grant PDR T.0120.21, with the participation of the Swiss National Science Fundation (SNF). MG. and EJ are F.R.S.-FNRS Research Directors.
The postdoctoral fellowship of KB is funded by F.R.S.-FNRS grant T.0109.20 and by the Francqui Foundation.
This publication benefits from the support of the French Community of Belgium in the context of the FRIA Doctoral Grant awarded to M.T.
GD acknowledges support by UKRI-STFC grants: ST/T003081/1 and ST/X001857/1.
PLP was partly funded by the FONDECYT Initiation Project No. 11241572.
 R.F.J. acknowledges support for this project provided by ANID's Millennium Science Initiative through grant ICN12\textunderscore 009, awarded to the Millennium Institute of Astrophysics (MAS), and by ANID's Basal project FB210003
JTR was partially supported by a CONICYT/FONDECYT 2018 Postdoctoral research grant, project number: 3180071 and acknowledges current financial support from both ANID/FONDECYT (2022 Initiation in Research grant; Project Number 11220287), and by the U.S. Air Force Office of Scientific Research (Award FA9550-22-1-0292).
LM acknowledges the financial contribution from the PRIN MUR 2022 project 2022J4H55R.
\section*{Data Availability}
Some of the light curves to derive mid-transit times were downloaded from the Exoplanet Transit Database at http://var2.astro.cz/ETD/. All other light curves appearing for the first time in this article are presented as online material. Mid-transit times derived from our own light curves, as well as those of other observers' and TESS light curves, are presented as online data sets too, through VizieR Online.



\bibliographystyle{mnras}
\bibliography{wasp4_manuscript_mnras_final} 




\appendix

\section{Mid-transit Times Used in this Study}
\label{app:midtransits}
We provide a few lines for the mid-transit timing data in five different tables below, the complete versions of which are presented as the online materials of this study.

\begin{table*}
\centering
\caption{Mid-transit times we derived from our own light curves.}
\begin{tabular}{ccccccccl}
 \hline \hline
 UT Date&Epoch&T$_{\rm c}$ (BJD$_{TDB}$) & $\sigma_{\rm T_c}$ (days) & Beta&PNR&Observatory & Elimination & Justification \\
\hline
\hline
2024/07/18&2992&2460509.737743 & 0.000383 &2.00 &1.37&El Sauce & 0 & -\\
2024/08/23&3019&2460545.868861 & 0.000232 &0.99 &2.94& TRAPPIST-South & 0 & -\\
2024/08/27&3022&2460549.883470 & 0.000742 &2.09 &2.89& El Sauce & 0 & -\\
2024/08/27&3022&2460549.883067 & 0.000234 &1.65 &2.98& TRAPPIST-South & 0 & -\\
2024/07/14&2989&2460505.722311 & 0.000876 &2.08 &1.84& El Sauce & 1 & Transit depth, Sigma clipping\\
2019/07/17&1626&2458681.712570 & 0.000228 &1.05 &1.24& La Silla & 0 & -\\
2019/07/25&1632&2458689.742452 & 0.000129 &1.30 &0.74& La Silla & 0 & -\\
... & ... & ... & ... & ...& ... & ...& ... & ... \\
\hline
\label{tab:Tc_ours}
\end{tabular}
\end{table*}

\begin{table*}
\centering
\caption{Mid-transit times we derived from TESS observations.}
\begin{tabular}{ccccccc}
 \hline \hline
 T$_{\rm c}$ (BJD$_{\rm TDB}$) & $\sigma_{\rm T_c}$ (days) & TESS Sector & PNR&Beta& Elimination & Justification \\
\hline
\hline
2458355.184997 & 0.000344 & 2 &1.79& 1.49& 0 & -\\
2458356.522484 & 0.000378 & 2 &1.96& 1.70& 0 & -\\
2458357.861048 & 0.000317 & 2 &1.76& 1.86& 0 & -\\
2458359.199613 & 0.000322 & 2 &1.61& 1.27& 0 & -\\
2458360.537197 & 0.000349 & 2 &1.86& 1.87& 0 & -\\
... & ... & ... & ... & ...& ... & ... \\
\hline
\label{tab:Tc_tess}
\end{tabular}
\end{table*}

\begin{table*}
\centering
\caption{Mid-transit times we derived from the light curves in the Exoplanet Transit Database (ETD) (T$_{\rm c}$, their uncertainties ($\sigma_{\rm T_c}$), mid-transit times reported by the observers of ETD from the same light curves (T$_{\rm rep}$), and their uncertainties ( $\sigma_{\rm T_{\rm rep}}$).}
\begin{tabular}{cccccclcl}
 \hline \hline
 T$_{\rm c}$ (BJD$_{\rm TDB}$) & $\sigma_{\rm T_c}$ (days) & T$_{\rm rep}$ (BJD$_{\rm TDB}$) & $\sigma_{\rm T_{\rm rep}}$ (days) &Beta&PNR& Observer & Elimination & Justification \\
\hline
\hline
2455385.649354 & 0.0002675717 & 2455385.649565 & 0.00027 &1.52&3.03& Sauer T. & 0 & -\\
2455425.796285 & 0.0001166734 & 2455425.796065 & 0.00011 &1.52&1.58& Fernández-Lajús, E. et al. & 0 & -\\
2455473.972411 & 0.0003775267 & 2455473.972356 & 0.00034 &1.72&1.86& Milne G. & 0 & -\\
2455506.090753 & 0.0004185778 & 2455506.090526 & 0.00044 &0.93&2.47&Curtis I. & 0 & -\\
2456556.602246 & 0.0001731441 & 2456556.602174 & 0.00019 &2.17&1.47& Villareal D. & 0 & -\\
... & ... & ... & ... & ... & ... & ...& ... & ...\\
\hline
\label{tab:Tc_etd}
\end{tabular}
\end{table*}

\begin{table*}
\centering
\caption{Mid-transit times we derived from literature light curves (T$_{\rm c}$, their uncertainties ($\sigma_{\rm T_c}$), mid-transit times reported in the literature from the same light curves (T$_{\rm rep}$), and their uncertainties ( $\sigma_{\rm T_{\rm rep}}$).}
\begin{tabular}{cccccclcl}
 \hline \hline
 T$_{\rm c}$ (BJD$_{\rm TDB}$) & $\sigma_{\rm T_c}$ (days) & T$_{\rm rep}$ (BJD$_{\rm TDB}$) & $\sigma_{\rm T_{\rm rep}}$ (days) &Beta&PNR& Reference & Elimination & Justification \\
\hline
\hline
2459411.049006 & 0.000166 & 2459411.048720 & 0.00016 &2.42&0.9& \citet{harre2023a} & 0 & -\\
2459436.475352 & 0.000137 & 2459436.475250 & 0.00015 &1.61&0.82& \citet{harre2023a} & 0 & -\\
2459444.504849 & 0.000158 & 2459444.504680 & 0.00028 &2.22&1.08& \citet{harre2023a} & 0 & -\\
- & - & 2459457.886750 & 0.00029 &-&-& \citet{harre2023a} & 1 & Beta criterion\\
2459465.916732 & 0.000194 & 2459465.916340 & 0.00029 &2.43&1.12& \citet{harre2023a} & 0 & -\\
2459480.636898 & 0.000161 & 2459480.636720 & 0.00022 &2.33&1.26& \citet{harre2023a} & 0 & -\\
2459502.048511 & 0.000154 & 2459502.048250 & 0.00026 &1.90&0.95& \citet{harre2023a} & 0 & -\\
... & ... & ... & ... & ... & ... & ...& ... & ...\\
\hline
\label{tab:Tc_lit1}
\end{tabular}
\end{table*}

\begin{table*}
\centering
\caption{Mid-transit times we made use of directly from the literature.}
\begin{tabular}{ccl}
 \hline \hline
T$_{\rm rep}$ (BJD$_{\rm TDB}$) & $\sigma_{\rm T_{\rm rep}}$ (days) & Reference \\
\hline
\hline
2454368.592790 & 0.00033 & \citep{wilson2008}\\
2454396.695760 & 0.00012 & \citep{gillon2009}\\
2454701.812800 & 0.00022 & \citep{hoyer2013}\\      
2454701.813030 & 0.00018 & \citep{hoyer2013}\\
2454705.827150 & 0.00029 & \citep{hoyer2013}\\      
... & ... & ... \\
\hline
\label{tab:Tc_lit2}
\end{tabular}
\end{table*}

\section{Transit Light Curves Acquired for this Study}
\label{app:lightcurves}
\begin{figure}
\includegraphics[width=\columnwidth]{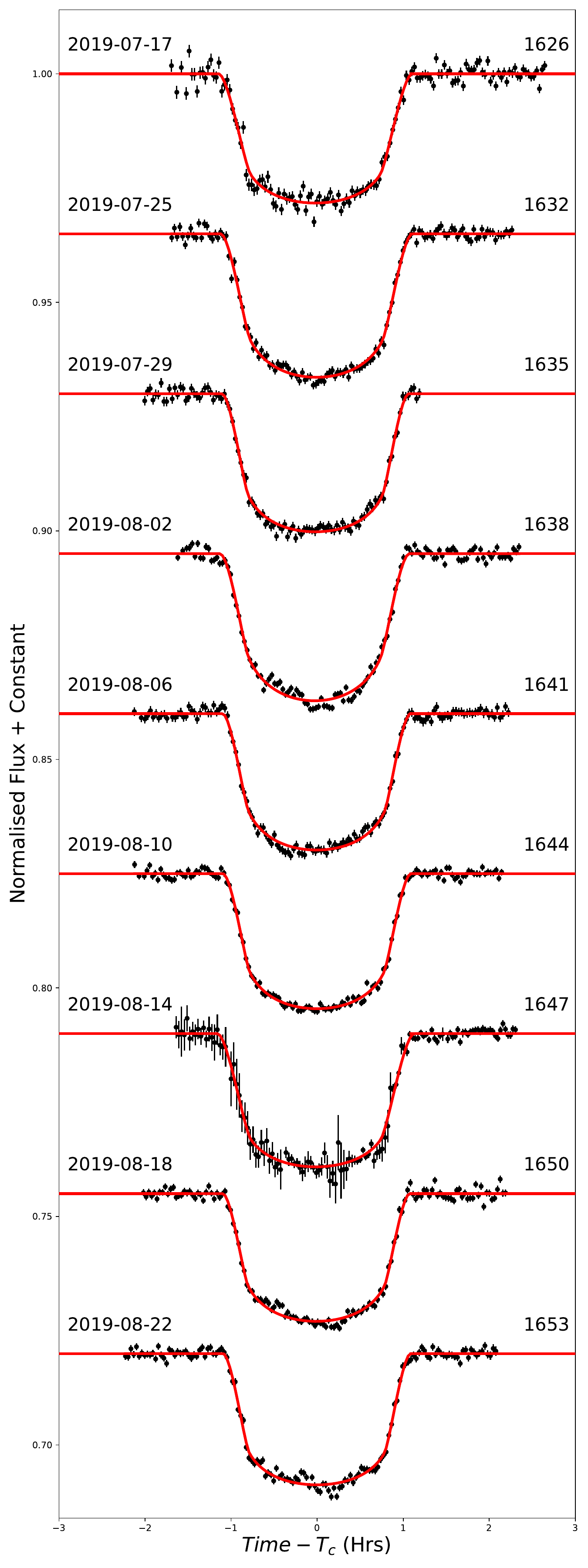}
\caption{Transit light curves we acquired in La Silla with the Danish Telescope (black dots with error bars) and their transit models (red continious lines). Epochs displayed at top right of each light curve. Eliminated light curves denoted with asteriks after UTC date.}
\label{fig:transits_01}
\end{figure}

\begin{figure}
\includegraphics[width=\columnwidth]{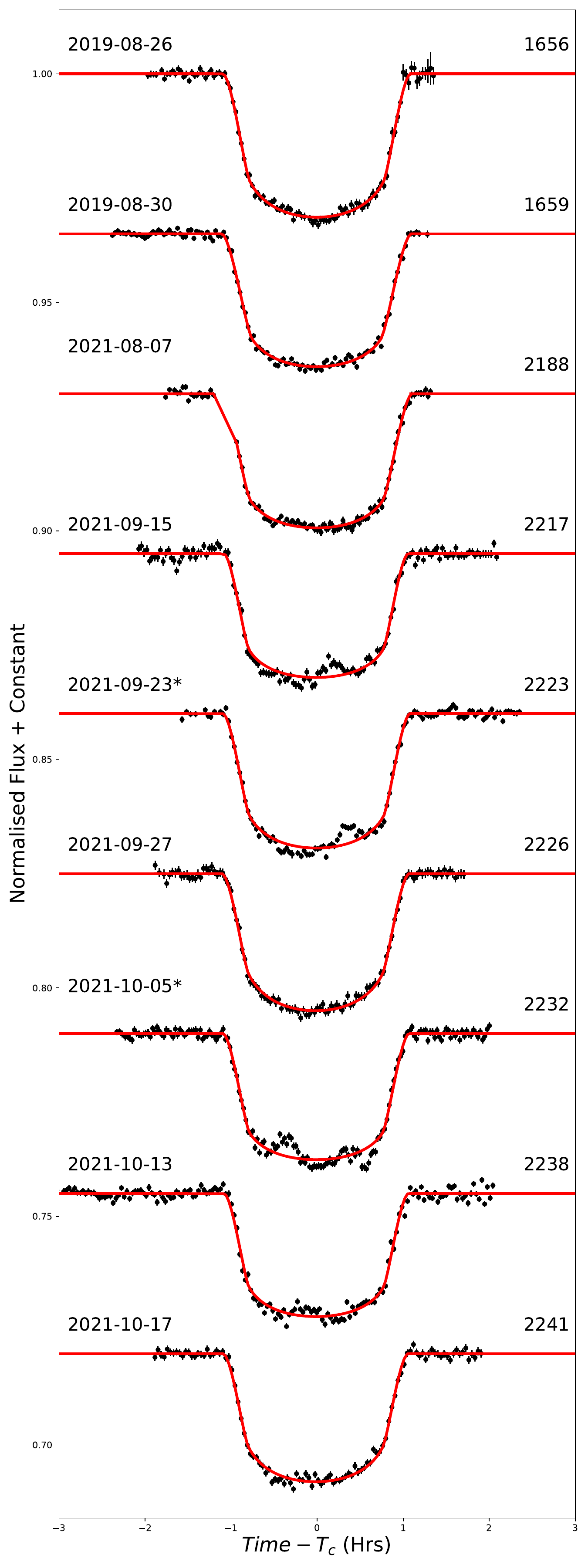}
\caption{Same as Fig. \ref{fig:transits_01}}
\label{fig:transits_02}
\end{figure}

\begin{figure}
\includegraphics[width=\columnwidth]{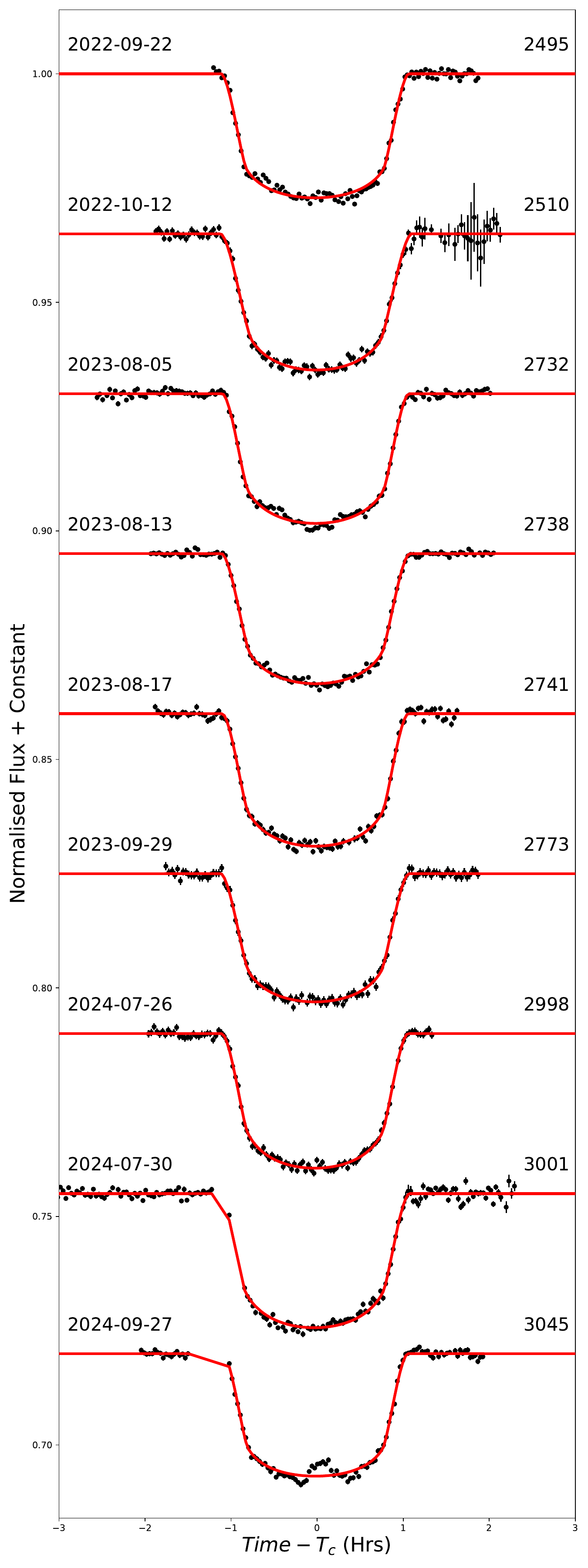}
\caption{Same as Fig. \ref{fig:transits_01}}
\label{fig:transits_03}
\end{figure}

\begin{figure}
\includegraphics[width=\columnwidth]{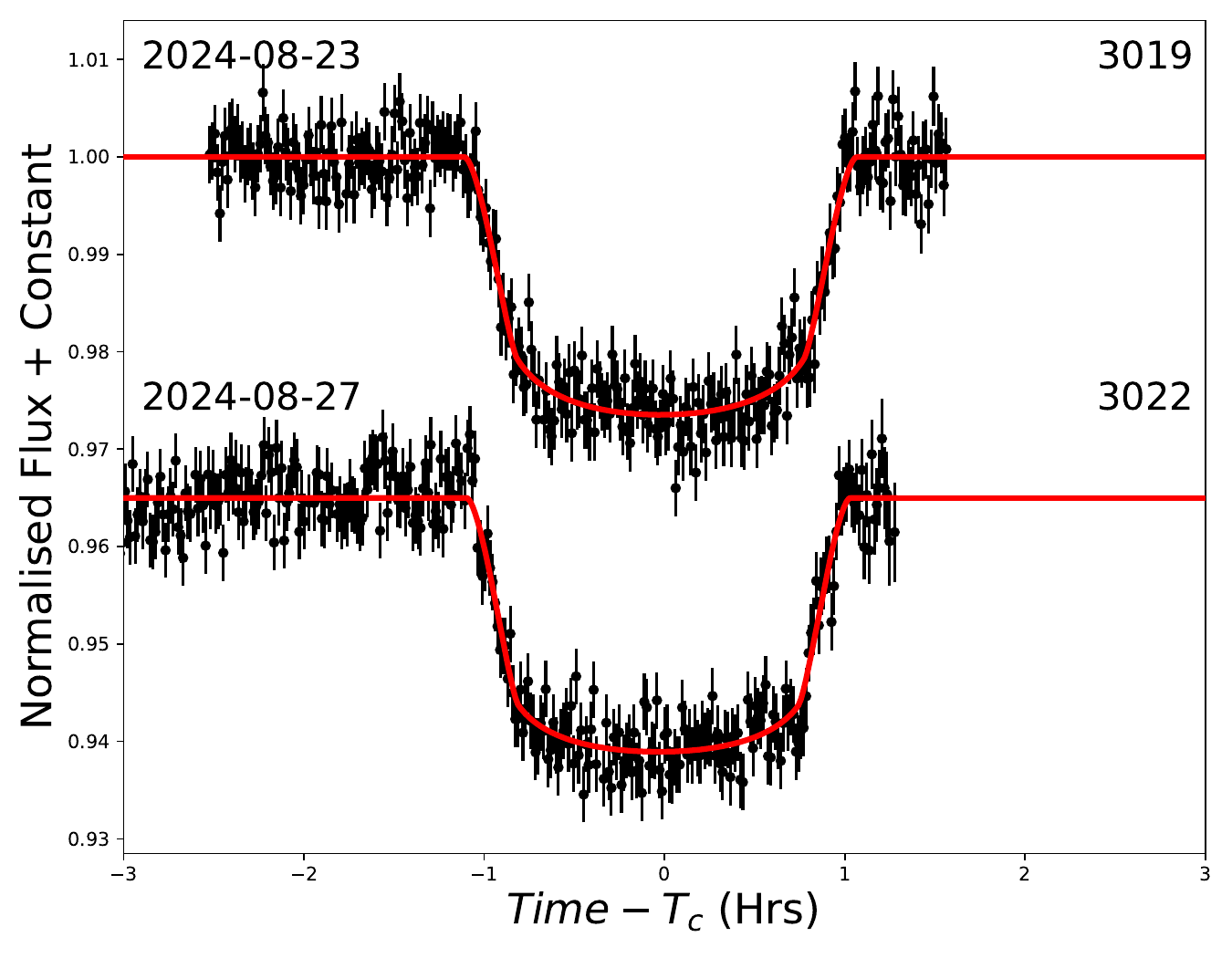}
\caption{Same as Fig. \ref{fig:transits_01} but for TRAPPIST observations.}
\label{fig:transits_05}
\end{figure}

\begin{figure}
\includegraphics[width=\columnwidth]{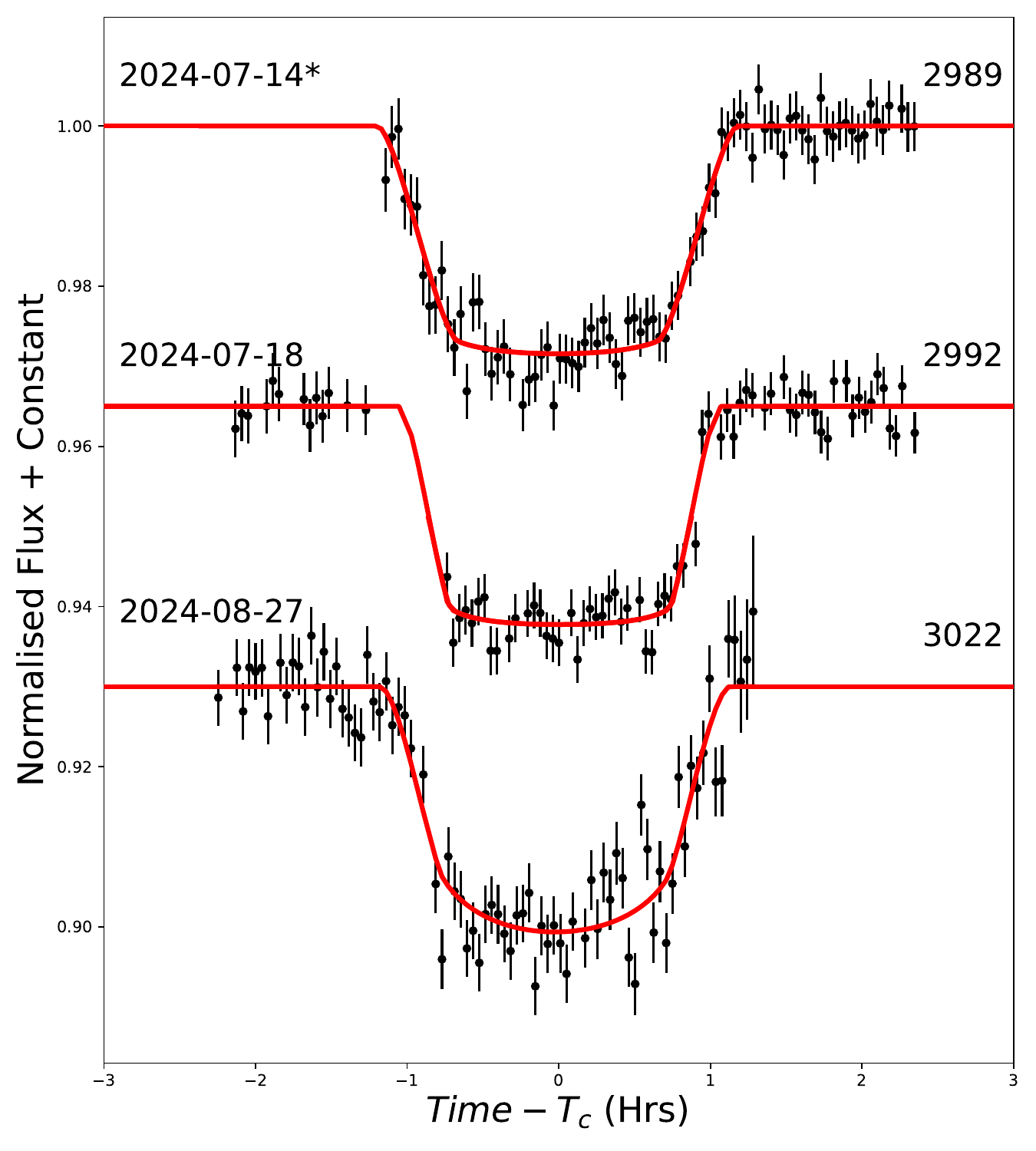}
\caption{Same as Fig. \ref{fig:transits_01} but for El Sauce observations.}
\label{fig:transits_04}
\end{figure}

\begin{figure}
\includegraphics[width=\columnwidth]{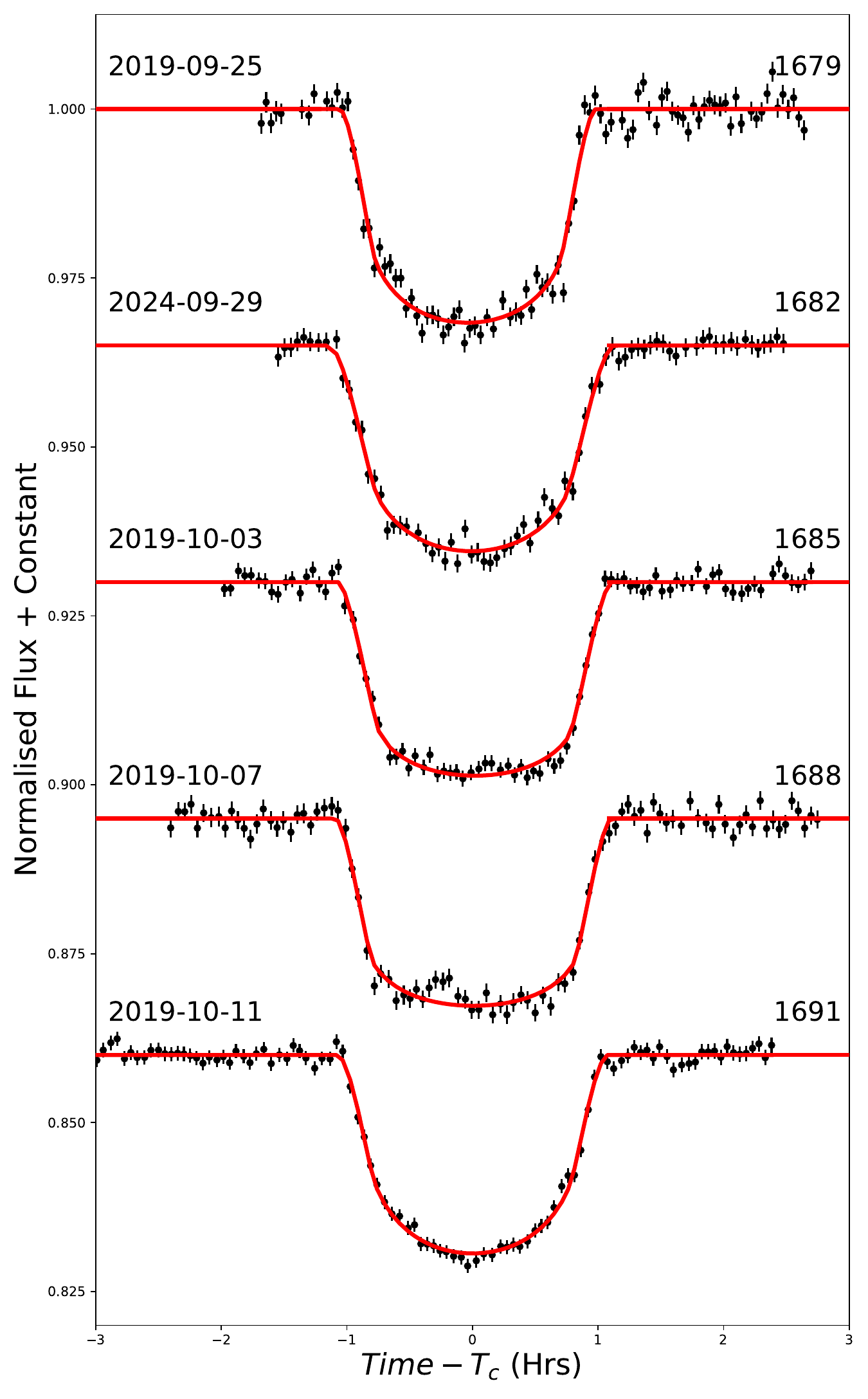}
\caption{Same as Fig. \ref{fig:transits_01} but for Ckoirama observations.}
\label{fig:transits_05}
\end{figure}



\bsp	
\label{lastpage}
\end{document}